\documentclass[twocolumn]{aastex631}
\usepackage{multirow}

\shorttitle{RV Orbit Refinement of Imaging Targets}
\shortauthors{Li et al.}

\begin{document}

\title{Radial Velocity Strategies for the Orbital Refinement of Exoplanet Direct Imaging Targets}

\newcommand{\num}{\textit{Num}}
\newcommand{\gap}{\textit{Gap}}
\newcommand{\cov}{\textit{Cov}}
\newcommand{\pmax}{\textit{$P_{max}$}}

\author[0000-0002-4860-7667]{Zhexing Li}
\affiliation{Department of Earth and Planetary Sciences, University of
  California, Riverside, CA 92521, USA}
\email{zli245@ucr.edu}

\author[0000-0002-7084-0529]{Stephen R. Kane}
\affiliation{Department of Earth and Planetary Sciences, University of
  California, Riverside, CA 92521, USA}

\author[0000-0002-3199-2888]{Sarah Blunt}
\affiliation{Department of Astronomy and Astrophysics, University of California, Santa Cruz, Santa Cruz, CA, 95064, USA}

\author[0000-0001-5737-1687]{Caleb K.\ Harada}
\altaffiliation{NSF Graduate Research Fellow}
\affiliation{Department of Astronomy, 501 Campbell Hall \#3411, University of California, Berkeley, CA 94720, USA}


\begin{abstract}

Many potential direct imaging candidates suffer from large orbital period uncertainties, leading to challenges in accurate predictions of future orbital positions and imprecise direct imaging measurements of planetary parameters. To improve the precision in orbital properties, precursor radial velocity (RV) follow-up observations for selected candidates are essential. This study examines the impact of three variables on the orbital period uncertainties of long-period giant planets: the number of future observations, the temporal gap between past and future data, and the temporal coverage of upcoming observations. Our simulations indicate that the orbital phases at which future RV observations are acquired play a significant role in reducing period uncertainties. Additionally, observing too frequently within a given time frame adds limited value to the program once a certain number of observations has been achieved. The temporal gap proves to be the most important factor when there is no strict end time to the observing campaign. However, if a strict end time is set, starting observations earlier yields improved reductions in orbital period uncertainty. These insights offer practical guidance for planning efficient RV follow-up campaigns to maximize the science yield of future space-based direct imaging missions.

\end{abstract}

\keywords{planetary systems -- techniques: radial velocity, direct imaging}


\section{Introduction}
\label{sec:intro}

One of the exoplanet community's primary goals is to directly image terrestrial planets within the habitable zone (HZ) of nearby stars, defined as the region around a star where the energy balance of the planet may enable the presence of surface liquid water \citep{kasting1993a,kane2012a,kopparapu2013a,kopparapu2014,kane2016c,hill2018,hill2023}. Many ongoing efforts are prioritizing the development of a next-generation space-based telescope capable of achieving this, such as the Habitable Worlds Observatory (HWO) recommended by the 2020 Astronomy and Astrophysics decadal survey \citep{nas2021}. To maximize the science yield of such missions, previous mission concept reports highlight the need for precursor science using alternative detection methods, including radial velocity (RV) observations, to provide the up-to-date information on the planetary systems of interest \citep{reportluvoir,gaudi2020}. Within the RV community, efforts are underway that push the detection limits of extreme precision radial velocity (EPRV) instruments \citep{fischer2016,jurgenson2016,schwab2016,pepe2021,gibson2024}. These RV/EPRV data may be applied to planning observations for imaging terrestrial planets \citep{newman2023}, understanding the orbital dynamics of known systems \citep{kane2024d,kane2024e}, addressing correlated noise in EPRV data \citep{luhn2023,gupta2024}, improving mass measurements through RV follow-ups \citep{burt2018,cloutier2018,meunier2023}, and constraining the presence of additional planets around target stars \citep{kane2019b,dalba2021,laliotis2023,harada2024b,harada2024c}. However, an area that remains underexplored is the extraction of precise orbital information for direct imaging (DI) candidates using RV data. While this may not pose significant issues for characterizing habitable terrestrial planets as these planets have typically been observed over numerous orbital periods, long-period giant planets in distant orbits, which will be key targets for the Nancy Grace Roman Space Telescope \citep[hereafter Roman;][]{spergel2015,akeson2019}, present unique challenges.

Roman is set to launch no later than May 2027 and will conduct a science demonstration mission highly relevant to the DI community. Equipped with the Coronagraph Instrument \citep[CGI;][]{riggs2021,bailey2023}, Roman will demonstrate several key technologies in space for the first time \citep{savransky2024}, which are essential for future DI missions such as the HWO in the search for habitable planets around nearby Sun-like stars. The performance and outcomes of Roman's CGI mission will have a significant impact on the design of the HWO, underscoring the importance of precursor science to support the success of this demonstration mission. Considering that Roman will likely be able to conduct deep imaging and spectroscopy for only a handful of Jovian analogs in reflected light \citep{wolff2024}, selecting the right candidates with the best-refined orbital parameters is critical to delivering high-quality science products for Roman.

Detecting a planet through DI requires it to be sufficiently bright and separated enough from its star to fall outside the instrumental inner working angle \citep{kane2013c,stark2014,brown2015a,kane2018c,li2021,vaughan2023}. These targets, ranging from distant gas giant planets to terrestrial planets in the HZ, will mostly have been observed by the RV method. However, previous work shows that, for long-period gas giants with years or decades-long orbits, the orbital uncertainty increases with the orbital period, mostly reaching around 10\% of the period value \citep{kane2007b,kane2009c}. In extreme cases, RV-derived orbital information can be nearly lost, as the uncertainty approaches the length of the orbital period itself \citep{kane2009c}, even for short-period transiting planets \citep{dragomir2020a,kane2021b}. Additionally, as pointed out in \citet{spohn2022}, existing RV solutions for planetary systems can grow ``stale" over time, losing their predictive accuracy for future DI observability. This presents a challenge for DI because planet--star flux ratios are sensitive to the planetary orbital positions, and prior knowledge about a planet's detectability is essential before DI observations \citep{turnbull2021}. Accurate extraction of planetary parameters from DI data relies on well-constrained orbital parameters, often derived from RV. Without precise orbital knowledge, there is a risk of scheduling observations at suboptimal epochs when the planet's brightness is reduced, or worse, when it falls below the instrument's contrast limit or within the inner working angle. Even when robust detections are possible despite orbital imprecision, large uncertainties in orbital periods can complicate estimations of planet--star separation, phase angle, and orbital phase, affecting albedo estimations and, ultimately, planetary characterization quality. However, valuable observing time on precision RV facilities is in high demand across many science programs, and the limited time allocated for RV programs supporting future DI missions is thus in urgent need of the most efficient observing planning \citep{harada2024c,kane2024e}. It is therefore critical to investigate how orbital ephemerides can be effectively refined such that precursor ground-based RV observations may be carried out most efficiently.

Previous works have attempted to investigate optimal RV scheduling using Fisher information \citep{baluev2008,baluev2009,cloutier2018,gupta2024,lam2024}. This method is mathematically elegant and presents a relatively easy and efficient way to estimate the performance of uncertainties without the need for heavy computation. However, conditions assumed by this analytical method may not hold in all cases and, when violated, could lead to an underestimation of errors or provide unreliable results. For example, Fisher information requires the linearity of equations of the RV model and parameters of interest, sufficient observations such that estimations are unbiased, and the final joint distribution can be estimated by a Gaussian profile \citep{baluev2008}. All of these conditions may not be satisfied in the case of cold gas giant planets that have relatively large orbital period uncertainties, short observational baselines compared to their respective orbital periods, sparse temporal sampling, and mild to high eccentricities. Numerical methods, despite the heavy computation required, do not require these conditions to be satisfied and can provide a more reliable error estimate. Because of these factors, we opted to conduct numerical simulations to estimate the uncertainties.

In this work, we present the results of a study on how the orbital period of distant giant planets can be effectively refined with respect to three observing variables: number of future observations (\num), temporal gap between past and future observations (\gap), and temporal coverage of future observations (\cov). It is intuitive to expect an inverse relationship between period uncertainty and these variables. Indeed, such relations were derived by \citet{lucy1971} for the number of total observations and the temporal baseline for all these observations, suggesting that more observations and a longer baseline are preferred. Here, instead of treating all observations as one dataset, the setup is slightly different, where we treat past and future data as two separate datasets and investigate only how these three variables collectively influence the orbital period uncertainty estimates. Specifically, we aim to include the mutual interactions among the three variables, examine the optimal observation timing for eccentric orbits, determine the conditions at which the uncertainty reduction reaches the point of diminishing return, and identify the relative importance of the variables in the model.

The paper is organized as follows: Section~\ref{sec:method} outlines the experiment setup, simulation variables, and validation steps we took to ensure robustness of our methods. Section~\ref{sec:sims} presents exploratory and a full suite of simulations on the influence of these variables on orbital period uncertainty. Short simulations using similar methods are presented in Section~\ref{sec:app} to demonstrate how these results could inform observing strategies. Finally, Section~\ref{sec:discuss} provides a discussion of the results of our simulations, and concluding remarks are contained in Section~\ref{sec:conclude}.


\section{Testing Scenarios and Synthetic RV Generation}
\label{sec:method}

    \subsection{Test System}
    
    For the purpose of this study, we focus on systems with at least one planet of interest to the DI community, particularly those with a distant orbit around their host star and relatively large orbital period uncertainties. The HD~134987 system is one such case and we used the original publication data as a test object. The discovery data spans $\sim$5000 days and comprises 138 observations acquired via the HIRES and UCLES spectrographs \citep{jones2010}. The system hosts two known planets, with orbital periods of $P_{b}$ = $259.19 \pm 0.07$ days and $P{c}$ = $5000 \pm 400$ days, for the b and c planets, respectively \citep{jones2010}. The system lies at a distance of 26.20~pc \citep{gaia2018} and the outermost planet (planet c) is one of the potential candidates for DI using Roman's CGI.

    \subsection{Test Scenarios}
    \label{sec:test}

    Given the relatively large orbital period uncertainty of planet c in the HD~134987 system, we aimed to conduct a suite of RV observing simulations to study the long-term behavior of errors associated with the best-estimated orbital period. The orbital period uncertainty behavior of distant planets was explored by generating synthetic RVs at future epochs and observing their long-term changes in uncertainty with respect to three variables: \num, \gap, and \cov. The synthetic RVs in this case are based on the best estimates of orbital parameters derived from modeling the original RVs provided by \citet{jones2010}. Since RV follow-up campaigns often have limited knowledge of the true orbital parameters of planets of interest, observations are typically acquired at a pre-determined cadence to maximize temporal coverage along the RV phase curve throughout the observing season. This approach assumes that the best estimates of orbital components, which may initially deviate from the true parameters, gradually converge toward the true values over time. Similarly, the best estimates used to generate synthetic future RVs for planet c carries an unknown offset from the true orbital parameters. To validate the reliability of using best-estimated parameters for synthetic RV generation and the robustness of the simulation outcomes, we conducted test runs on the three variables mentioned above, analyzing each across three independent synthetic data creation methods:
    
    \begin{enumerate}
        \item Future RVs are calculated using the true parameters and the RVs are sampled at epochs according to those parameters. This represents an ideal (though unlikely) scenario in which the true parameters are known.
        \item Future RVs are calculated using the true parameters, but the sampling of RVs is based on the best-estimate parameters. This reflects a realistic scenario for RV follow-up campaigns, where observational planning relies on the current best estimates, even though RVs are ultimately generated from a ground-truth source.
        \item Future RVs are calculated using best-estimate parameters, and the sampling of RVs is also based on those estimates. This scenario represents our simulation setup, where both RV generation and sampling depend on the estimated values.
    \end{enumerate}

    In each scenario, test runs were conducted by varying one variable at a time while keeping the other two constant. We adjusted the \num~variable from 6 to 200 in steps of 1, with \gap~and \cov~fixed at the full period of the outermost planet (\pmax); The \gap~variable, measured in units of \pmax, was varied from 0.1 to 20$\times$\pmax in steps of 0.1, while \num~was fixed to 50 and \cov~to 1$\times$\pmax. The same range and step size were applied to \cov, with \num~and \gap~set to 50 and 1$\times$\pmax, respectively. Future RV sampling epochs were determined by these three variables. For example, if the last archival RV observation occurred on 2,455,500 BJD in a system where \pmax~has a true orbital period of 1000 days and a best-estimate period of 1500 days, with the three variables set to 50, 1$\times$\pmax, and 1$\times$\pmax, the starting times for future RVs would be 2,455,500 + 1$\times$1000 = 2,456,500 BJD for Scenario 1, and 2,455,500 + 1$\times$1500 = 2,457,000 BJD for Scenario 2 and 3 (with \gap~set to 1$\times$\pmax, true values for Scenario 1 and estimated values for Scenario 2 and 3). The corresponding ending times would be 2,456,500 + 1$\times$ 1000 = 2,457,500 BJD for Scenario 1, and 2,457,000 + 1$\times$1500 = 2,458,500 BJD for Scenario 2 and 3 (with \cov~set to 1$\times$\pmax). With the start and end times defined, synthetic RVs were populated roughly evenly between these timestamps, with some randomness introduced (as described below). Scenario 1 represents the most ideal, or ``ground truth" case, where the results from this scenario are used as a control case against which results from other scenarios can be compared. Scenario 2 uses true parameters but bases its sampling on the best-estimate values, leading to minor deviations in sampling phases from the ground truth, therefore less optimal. Scenario 3 is the least ideal case, using only best-estimate parameters for both RV generation and sampling, and reflects the practical limitations due to estimation error. Our main interest lies in whether the orbital period and its uncertainty behavior observed in Scenario 3 can approximate those in Scenario 1.
    
    For the test runs on HD 134987, we adopted the orbital parameters derived from our modeling of the original RV data in \citet{jones2010} as the ground truth, where \pmax~=~5227 days. Again, these parameters are generally unknown to observers. Using these parameters, we generated synthetic RVs with the same measurement errors and RV scatter as the original RVs. The modeling of the synthetic RVs produced the best-estimate parameters for the system, where \pmax~=~5333 days. Given a set of values for the three variables, as well as the orbital parameters (true or estimated, depending on the scenarios), future synthetic RV timestamps were then evenly populated between the calculated starting and ending times. The number of timestamps created was set to be 50\% more than the intended, with the additional RVs later removed. These timestamps were then passed through a Gaussian filter with a sigma of 10 days to further increase the variation among created times. The synthetic RV at each timestamp was then calculated for each planet in the system by solving Kepler's equations using either true or estimated orbital parameters (again depending on the scenarios). RVs were calculated using Equation (\ref{eqn:RV}), where $V_{0}$ is the RV offset (zero in this case), while $\omega$, $f$, $K$, and $e$ are argument of periastron, true anomaly, semiamplitude, and orbital eccentricity of the planet, respectively. The total RVs for the system were simply the superposition of the individual velocity contributions from each planet. These total RVs were then passed through another Gaussian filter with a sigma equal to the rms of the fit to the original \citet{jones2010} data to mimic the combined noise from stellar and instrumental sources. Finally, the additional 50\% of the RVs introduced earlier were removed from the synthetic time series at randomly selected timestamps to mimic uneven temporal sampling. We note that this process ignores many important factors that are crucial for RV observations, such as correlated noise and the seasonal visibility. These are discussed in Section \ref{sec:discuss}. In short, factors like these are highly target dependent. In order to provide an unbiased and uniform approach for any target, we opted to not include these factors and therefore our results will likely represent a lower limit on the uncertainty.

    \begin{equation}
        \label{eqn:RV}
        V = V_{0} + K[\cos (\omega +f) + e \cos \omega ]
    \end{equation}

    The synthetic RVs creation process was conducted at each iteration of each variable under each scenario within our automated pipeline. At each iteration, the generated dataset was fed into the RV modeling toolkit {\sc RadVel}, which employs maximum a posteriori optimization for RV fitting and Markov Chain Monte Carlo (MCMC) for confidence interval estimation. Further details about {\sc RadVel} can be found in \citet{fulton2018a} and on its documentation page. Our automated pipeline logs each step of the MCMC process and saves the final derived Keplerian orbital parameters along with their associated error estimates from the MCMC posteriors in separate files. Depending on the characteristics of the system and the available RVs, the errors estimated from the posterior distributions can be large if the system is poorly constrained. In such cases, results from each MCMC run may vary due to the probabilistic nature of the MCMC process and the broader peaks in the posteriors. To minimize the impact of any single run skewing the result, we repeated the {\sc RadVel} fitting process five times per iteration and then averaged the results from these five MCMC runs to obtain a final set of derived parameters and associated errors. The final error used to gauge the orbital period uncertainty performance was simply the average of its upper and lower bound error. 
    
\begin{figure*}[tbp]
        \begin{center}
            \begin{tabular}{cc}
                \includegraphics[trim=0 0 0 0,clip,width=0.5\textwidth]{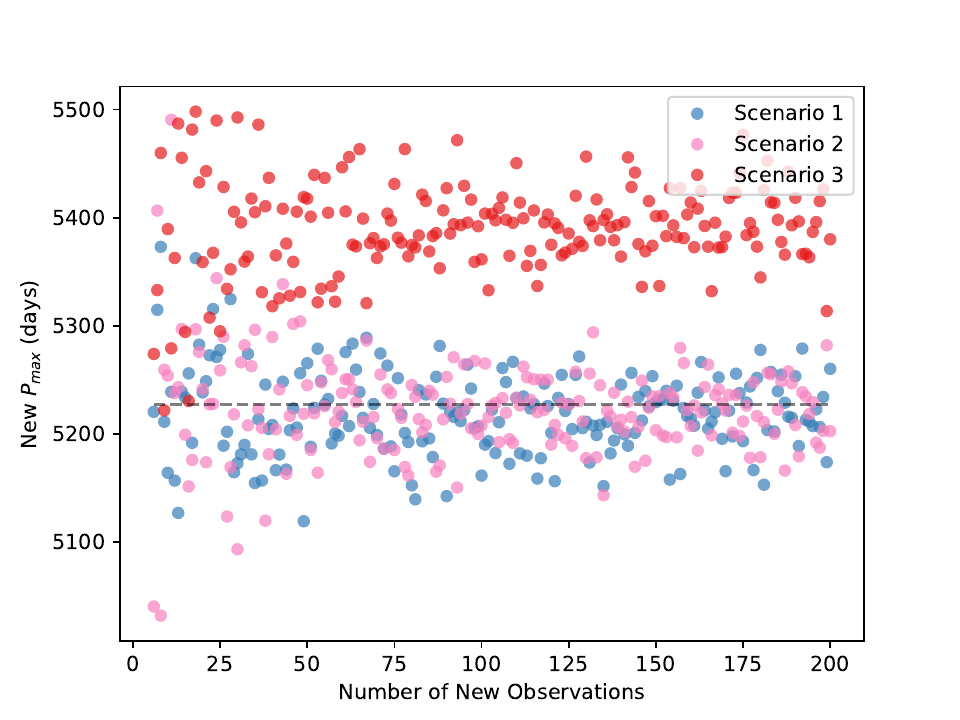} &
                \includegraphics[trim=0 0 0 0,clip,width=0.5\textwidth]{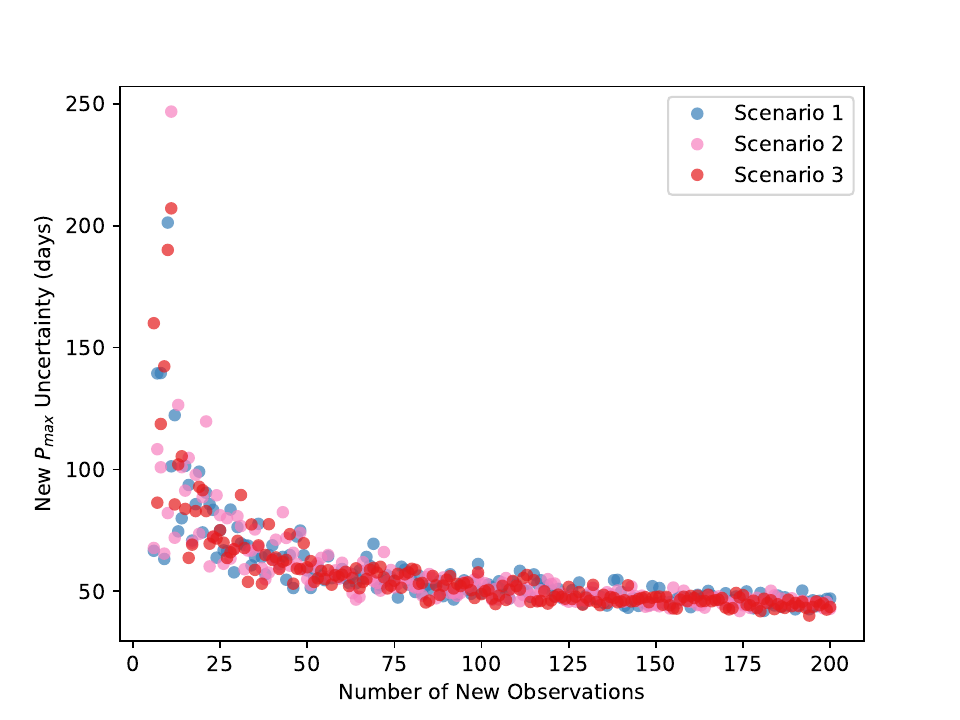} \\
                \includegraphics[trim=0 0 0 0,clip,width=0.5\textwidth]{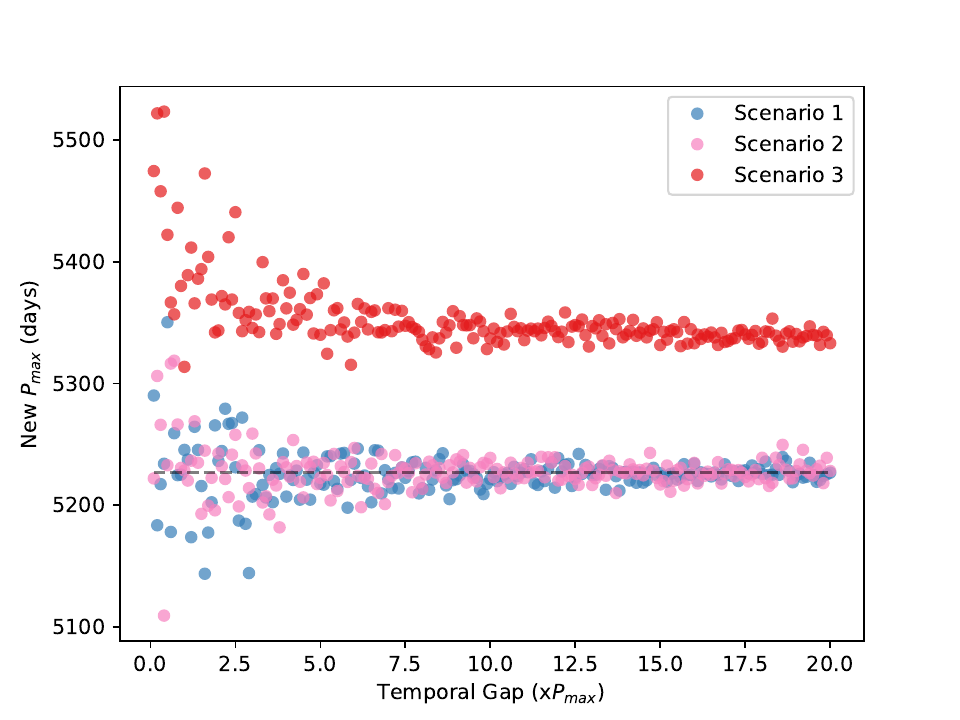} &
                \includegraphics[trim=0 0 0 0,clip,width=0.5\textwidth]{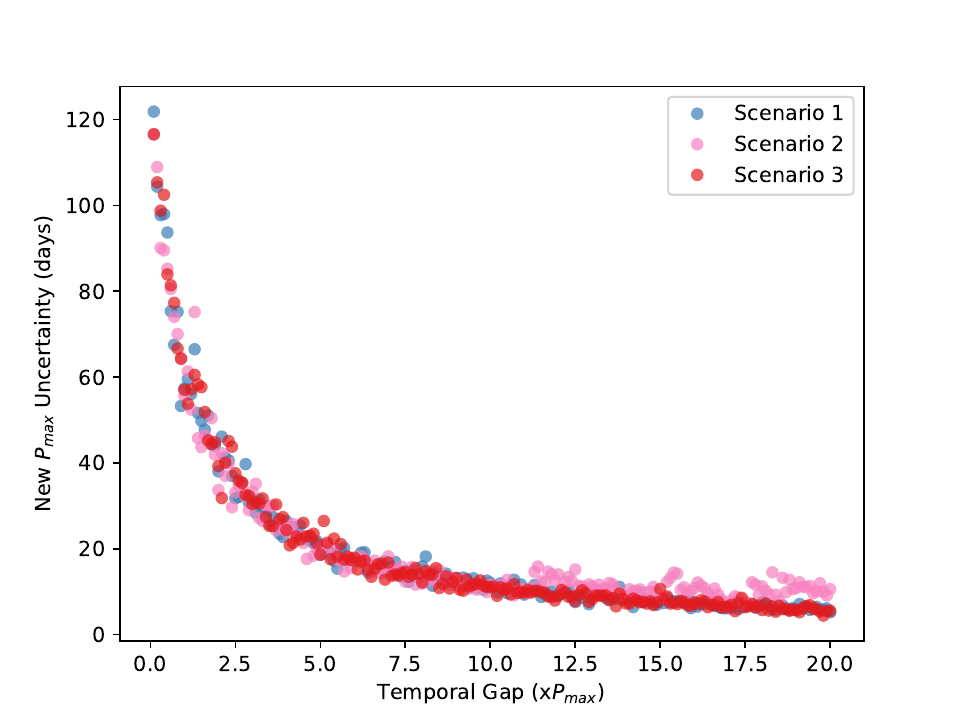} \\
                \includegraphics[trim=0 0 0 0,clip,width=0.5\textwidth]{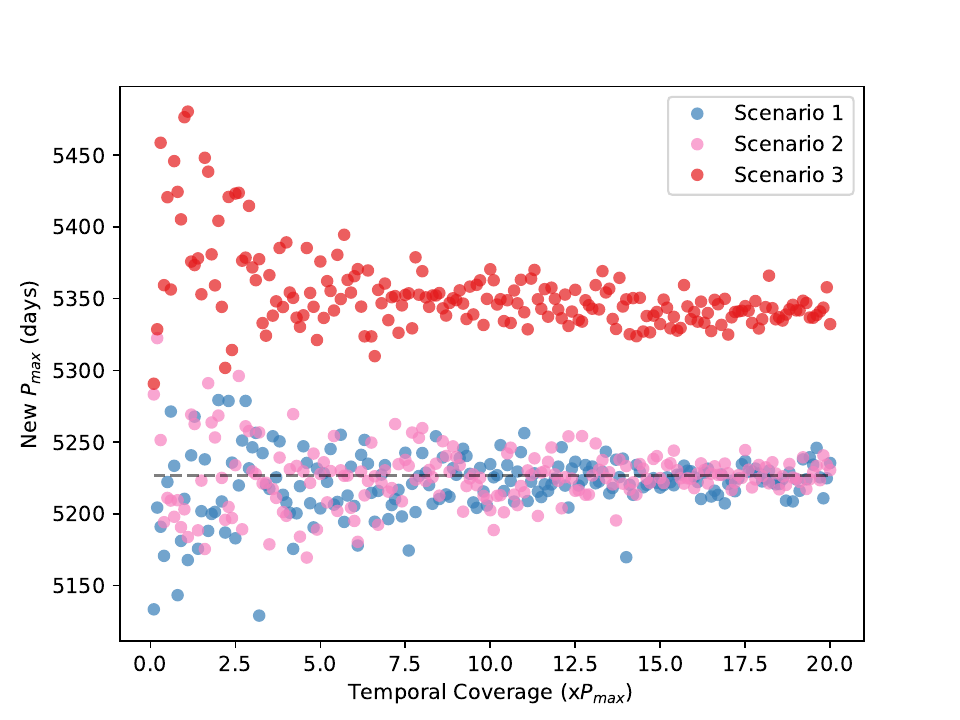} &
                \includegraphics[trim=0 0 0 0,clip,width=0.5\textwidth]{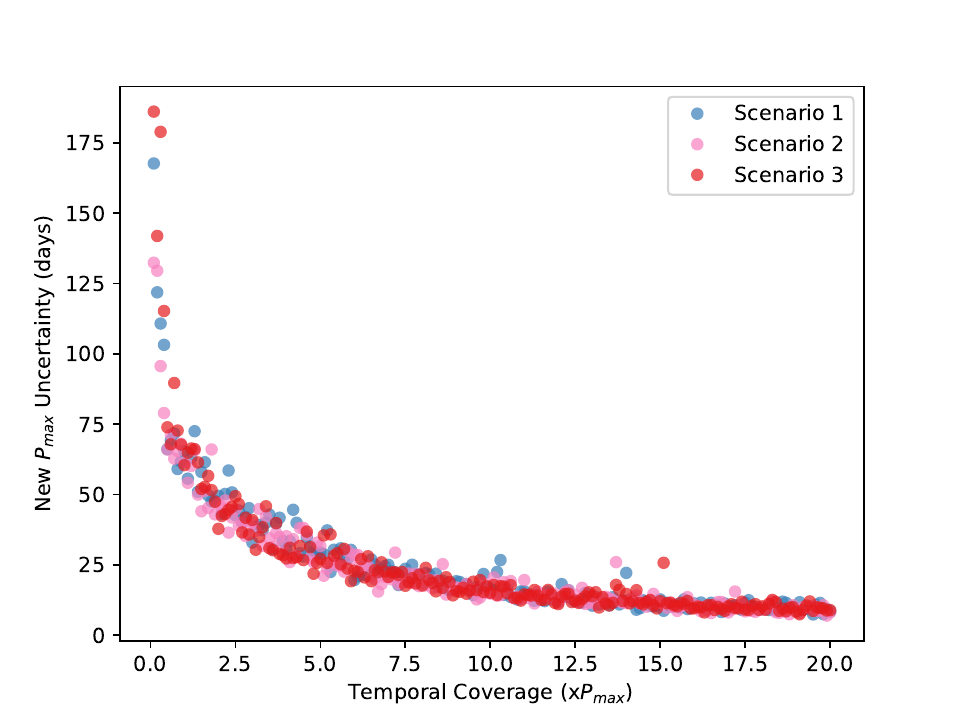}
            \end{tabular}
        \end{center}
        \caption{Testing result for orbital period and uncertainty evolution of the HD 134987 c planet based on synthetic RVs. Three scenarios, or data creation methods, are color coded in each panel. The left column shows orbital period and the right column shows associated uncertainty. The horizontal dashed line in the left panels represent the presumed ground truth period mentioned in the text. Top, middle, and bottom panels are results for the \num, \gap, and \cov~variables, respectively.}
    \label{fig:testruns}
    \end{figure*}

    In Figure \ref{fig:testruns}, we present the results from the testing simulations. Each row contains results for one variable, with the left column showing the behavior of the orbital period and and the right column displaying the associated uncertainty for each variable. In the left column across all three variables, the orbital periods in Scenario 1 and 2 both converge to the true value at a similar rate over time, as expected, given that the synthetic RVs were generated using the ``ground truth" values despite minor differences in the starting and ending times due to offsets in the true and estimated period values. For Scenario 3, the periods exhibit similar behavior, converging with comparable scatter to the other two scenarios as \num, \gap, or \cov~variables increase. However, in this case, the periods converge to the estimated orbital period used for generating the synthetic future RVs, which aligns with expectations. In the right column, it is notable that, despite the different methods of generating and sampling future RVs, the uncertainties for \pmax~follow nearly identical trajectories across all three variables: a rapid initial decrease in error estimates that gradually levels off. This consistent behavior indicates that, even with the less optimal RV creation and sampling approach of Scenario 3, we were able to replicate the long-term uncertainty behavior observed in real RV data. Consequently, we adopt the approach of Scenario 3 to create and sample RVs in all subsequent simulations.

\section{Simulations}
\label{sec:sims}

    \subsection{Exploratory Simulations}
    \label{sec:exploresims}

\begin{figure*}[tbp]
        \begin{center}
            \begin{tabular}{cc}
                \includegraphics[trim=0 0 0 0,clip,width=0.48\textwidth]{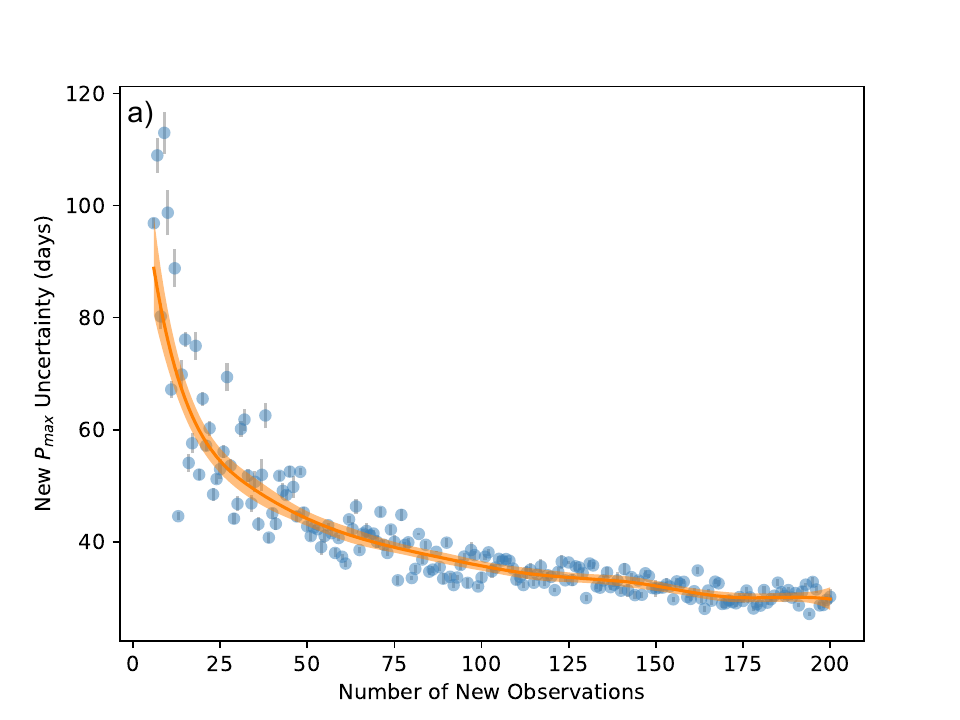} &
                \includegraphics[trim=0 0 0 0,clip,width=0.48\textwidth]{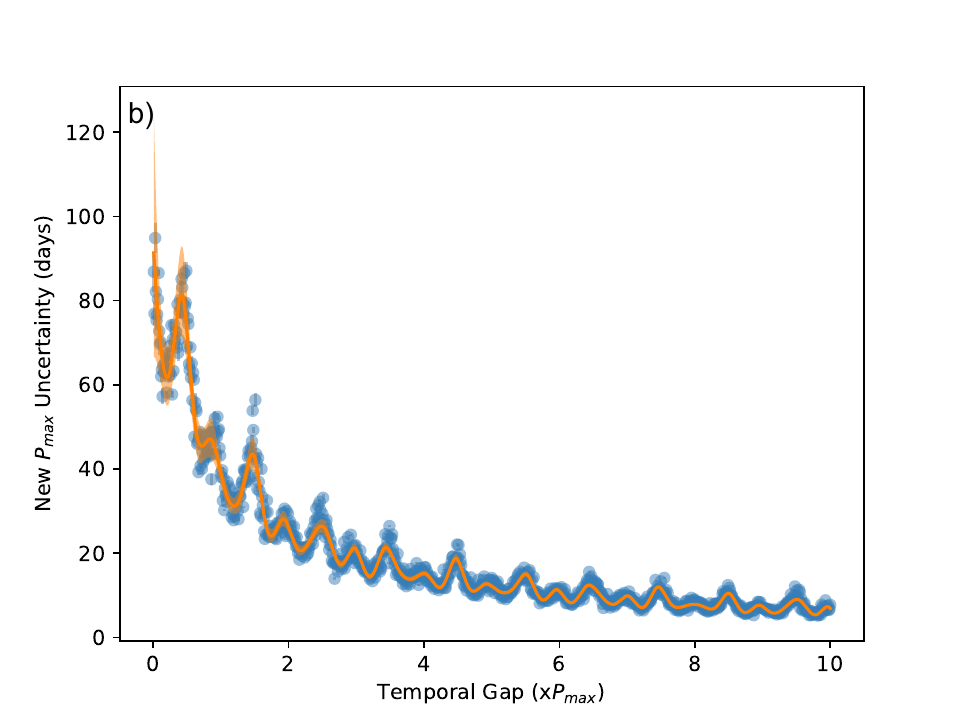} \\
                \includegraphics[trim=0 0 0 0,clip,width=0.48\textwidth]{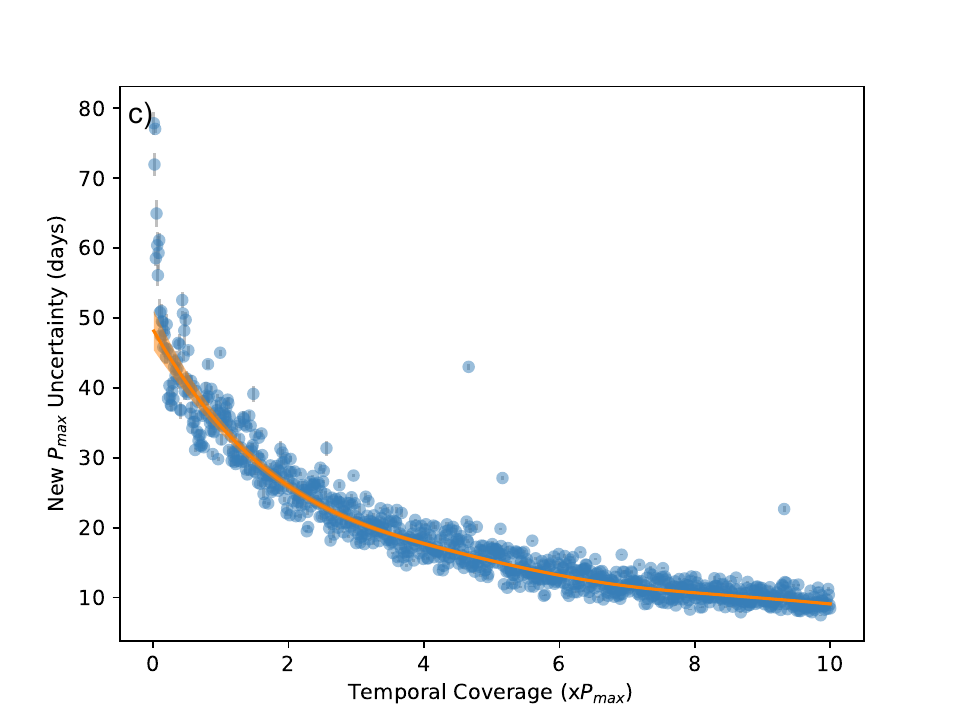} &
                \includegraphics[trim=0 0 0 0,clip,width=0.48\textwidth]{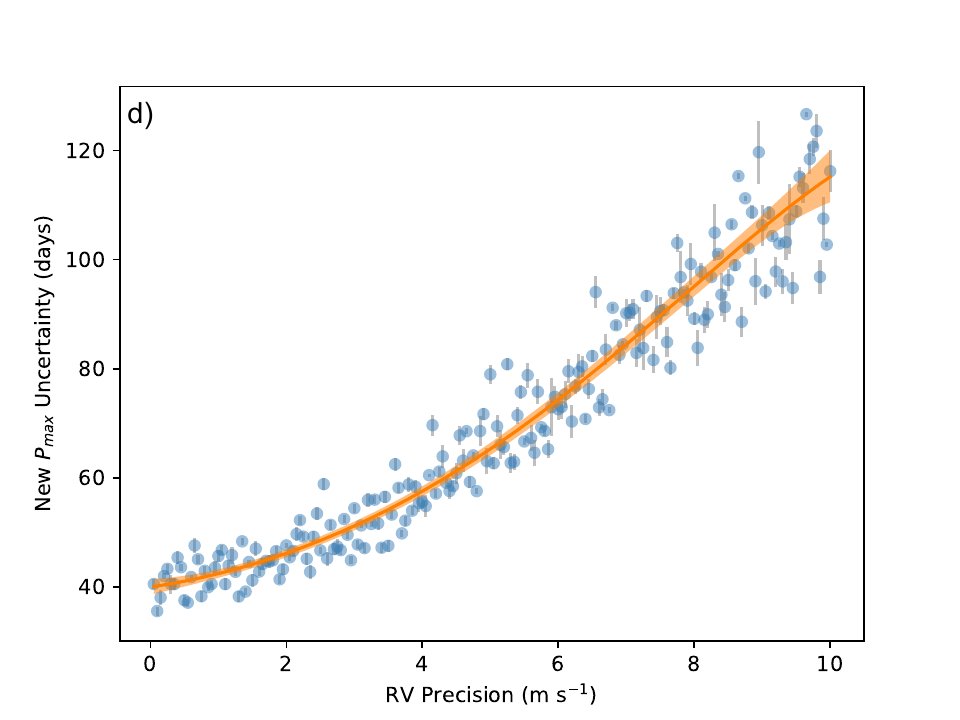}
            \end{tabular}
        \end{center}
        \caption{Simulation results from HD 134987 exploratory runs with synthetic data created at future epochs using four variables and real data from \citet{jones2010}. The \pmax~uncertainty before the simulations is 410 days. From panel (a) to panel (d) are \pmax~uncertainty with respect to: \num, \gap, \cov, and $Inst$ variables, respectively. Raw data are in blue and the best fit model with confidence interval for each variable is in orange. Uncertainty humps and dips present in the (b) panel was due to the fact that sampling at certain future epochs/RV phases are preferred for orbital period uncertainty reduction.}
    \label{fig:exploreruns}
    \end{figure*}

    Having selected the strategy for generating synthetic RVs, we proceeded with a set of exploratory simulations on the HD~134987 system. The procedure closely followed the testing process described in Section~\ref{sec:test}, with the exception that we adopted the Scenario 3 strategy and used real archival RVs from the original RV dataset provided by \citet{jones2010}. The \pmax~uncertainty behavior was tested in response to changes in the variables, \num, \gap, and \cov. Additionally, we introduced a fourth variable, instrumental precision ($Inst$), to examine its impact on the uncertainty of \pmax. We conducted four exploratory runs, each varying one of these variables while holding the others constant: \num~was fixed at 50, \gap~at 0.75$\times$\pmax, and \cov~at 0.35$\times$\pmax, except for their respective variable-specific runs. Instrumental precision was set to the average RV error from \citet{jones2010}, which is 1.26 m~s$^{-1}$, except in the fourth run. Specifically, in run 1, \num~was varied from 6 to 200 with a step of 1; in run 2, \gap~was varied from 0.01$\times$\pmax~to 10$\times$\pmax with a step of 0.01$\times$\pmax; in run 3, \cov~was varied from 0.01$\times$\pmax~to 10$\times$\pmax with a step of 0.01$\times$\pmax; and in run 4, $Inst$ ranged from 0.05 m~s$^{-1}$ to 10 m~s$^{-1}$ with a step of 0.05 m~s$^{-1}$. As with the previous testing runs, the modeling process was repeated five times at each iteration for each variable to average out variations from the MCMC output. Results were then modeled and the method used for modeling the best-fit curve is described in Section~\ref{sec:app}. Both the raw results and best fits for the HD 134987 exploratory runs are shown in Figure~\ref{fig:exploreruns}.

    For comparison, the orbital period uncertainty for the HD 134987 c planet from our modeling of the \citet{jones2010} data was 410 days. Results from the exploratory runs indicate that the orbital period uncertainty of this outer planet followed a similar uncertainty reduction pattern across all the three variables: \num, \gap, and \cov, as in the testing runs from the previous section. In all three cases, uncertainty dropped significantly initially, with a slower rate of decrease as the magnitude of each variable increased. The uncertainty behavior of each of the individual variables is expected since \citet{lucy1971} pointed out the inverse relationship between period error estimates and both number of observations and temporal baseline. For the $Inst$ variable, it is unsurprising that orbital period uncertainty increased with larger instrumental precision at future epochs. However, the improvement in precision from meter-per-second instruments to extreme-precision instruments such as those discussed in \citet{crass2021} was minimal in terms of uncertainty reduction. This suggests that, for reducing orbital period uncertainty of distant gas giant planets, there is little gain from observations using EPRV instruments and therefore utilizing the meter-per-second facilities will suffice for this task. For this reason, we do not explore instrumental precision as another variable and instead assume that all our future RVs would have a similar precision level to past data.
    
\begin{figure*}[tbp]
        \begin{center}
            \begin{tabular}{cc}
                \includegraphics[trim=0 0 0 0,clip,width=0.48\textwidth]{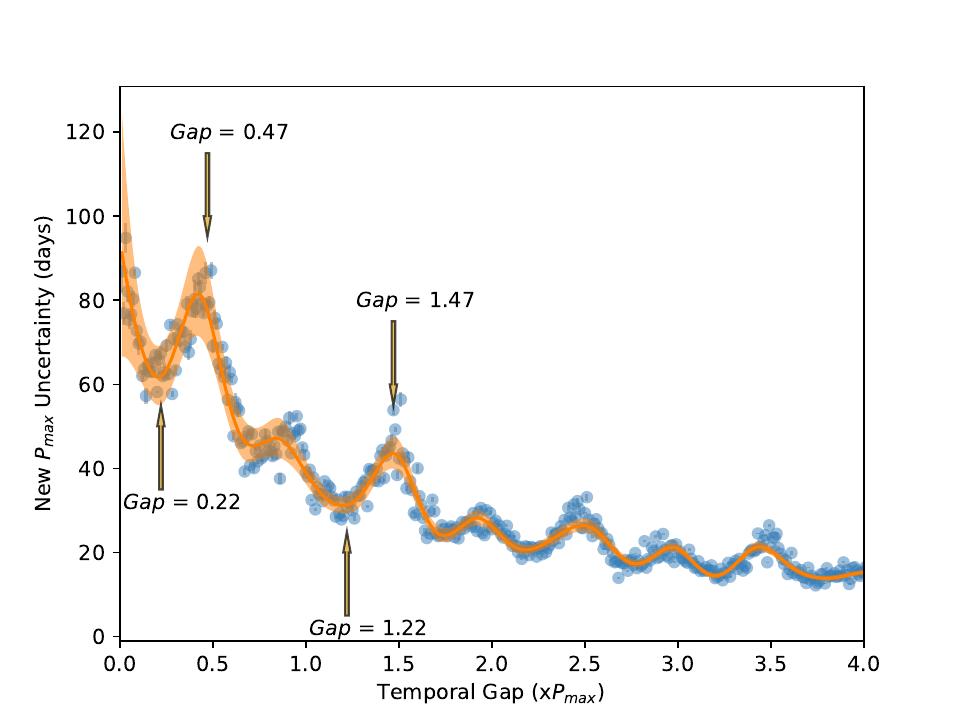} &
                \includegraphics[trim=0 0 0 0,clip,width=0.48\textwidth]{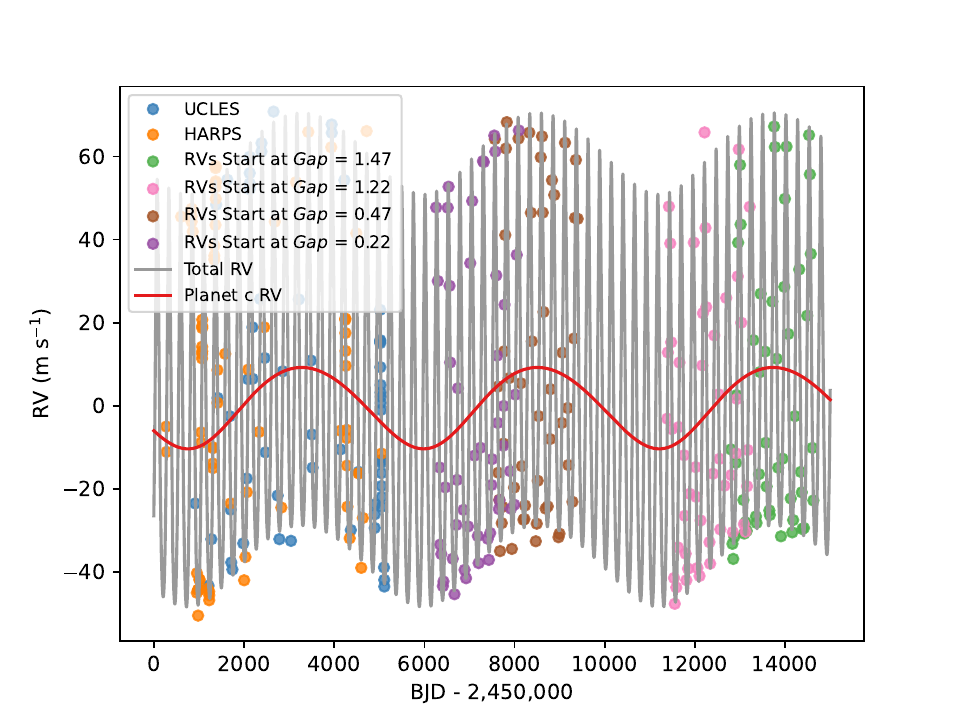} \\
                \includegraphics[trim=0 0 0 0,clip,width=0.48\textwidth]{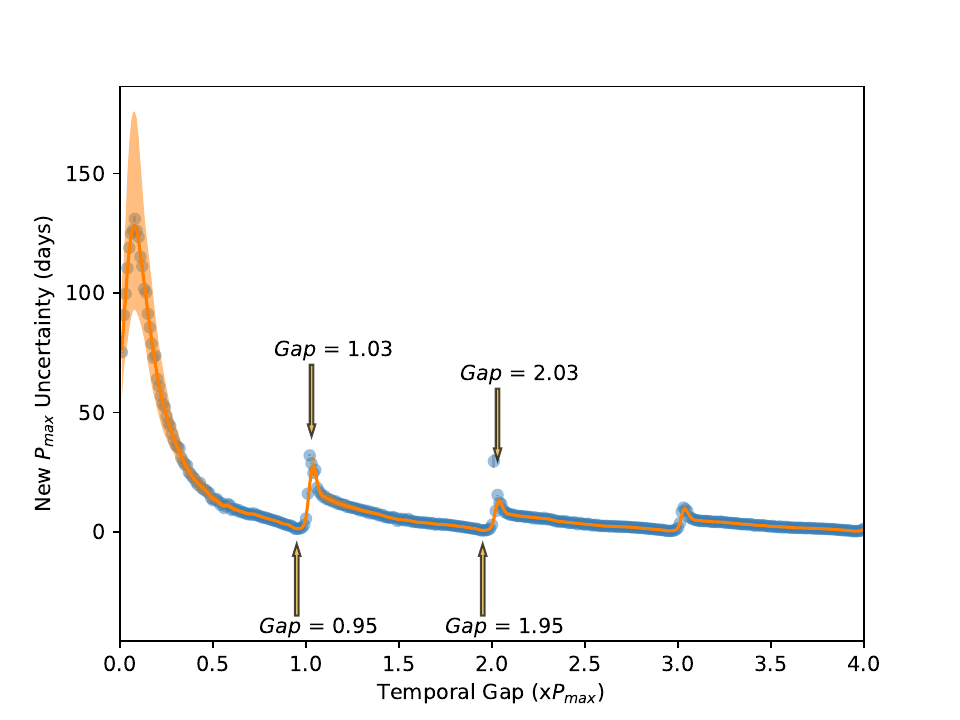} &
                \includegraphics[trim=0 0 0 0,clip,width=0.48\textwidth]{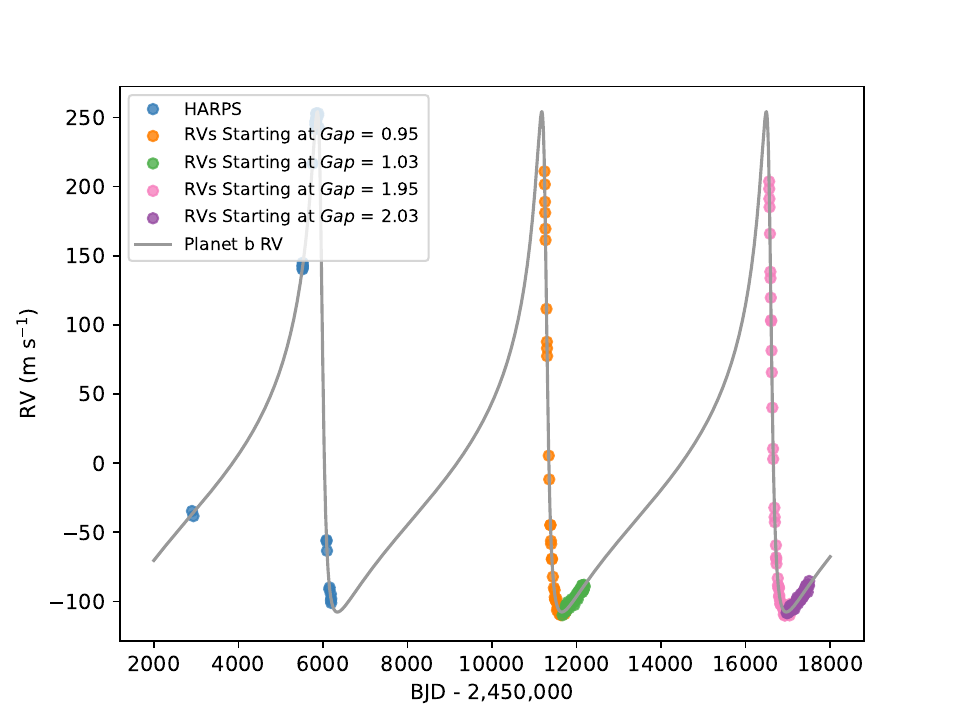}
            \end{tabular}
        \end{center}
        \caption{Simulation results of HD~134987 (upper panels) and HD~219077 (lower panels) for the \gap~variable. Left panels show the \pmax~uncertainty results from the simulation. Runs highlighted by the arrows near the humps and dips in the left panel corresponds to runs displayed in the right panel. Archival data, synthetic data starting at different \gap~values, total RV, and the outer most planet's RV are shown in the right panels. Synthetic RVs were populated according to the orbit of the outer most planet. The \pmax~uncertainties before the simulations for HD~134987 and HD~219077 are 410 and 140 days, respectively}
    \label{fig:case2}
    \end{figure*}

    One noticeable difference here is that the uncertainty reduction curve for the \gap~variable shows periodic humps and dips. This occurs because, in the exploratory runs, the \cov~variable was set to only 0.35 times the orbital period of planet c. As a result, at each iteration, the generated synthetic RVs at future epochs covered only a portion of planet c's orbital phases. In contrast, for the testing cases in the previous section, \cov~was set to 1$\times$\pmax and covered the entire period space and therefore avoided any bias in the sampled RV phases. Depending on the \gap~variable provided, the generated data may cover either the maximum or minimum phases of planet c's RV curve or, in some cases, the zero-crossing phases. In the case of HD 134987, humps and dips appear periodically with alternating amplitudes: one large hump and dip followed by a smaller one within each temporal gap interval. Initially, one might expect such a hump and dip pair to occur when future RVs are observed near zero-crossing and maximum/minimum phases of the RV curve, with the new data providing weaker and stronger constraints on the orbital parameters, respectively. However, upon closer examination of the exploratory simulation data, we observe that orbital period uncertainty reduction appears to be greatest when future RVs were populated near the zero-crossing phases rather than near the maximum/minimum phases of the planet c's RV curve. In Figure~\ref{fig:case2}, we highlight two groups of humps and dips, separated by one period. The upper panels represent runs for HD 134987, with the left upper panel resembling the \gap~variable result from Figure~\ref{fig:exploreruns} but with a shortened \gap~range on the horizontal axis to focus on the first few groups of features. Specific \gap~values at which future data begins are indicated by arrows in the left panel. In the upper right panel, both the total RV and planet c's contribution, along with archival and synthetic RVs are shown. Here, \pmax~uncertainty humps at \gap~=~0.47$\times$\pmax~and \gap~=~1.47$\times$\pmax in the upper left panel align with phases near the RV maximum for planet c, while synthetic RVs populated near zero-crossing phases correspond to uncertainty dips at \gap~=~0.22$\times$\pmax~and \gap~=~1.22$\times$\pmax. This suggests that, for optimal orbital period constraint, future observations should target phases where the gradient of the RV curve is steepest; that is, when the rate of RV change over time is greatest.
    
    To verify this, we conducted another simulation for the \gap~variable on HD~219077, a system with a single known planet exhibiting an extremely high eccentricity of 0.769 and an argument of periastron of 57 degrees for the host star \citep{marmier2013}. The setup was similar to that for HD~134987, except we only used HARPS data and limited the \gap~variable to a range of 0.01$\times$\pmax~to 4$\times$\pmax, with a step of 0.01$\times$\pmax. The \cov~variable was set to only 0.01$\times$\pmax to make sure the simulation data was able to capture the fast and short RV turnaround. The results are shown in the bottom panels of Figure~\ref{fig:case2}. The bottom left panel shows a similar decline in \pmax~uncertainty with periodic humps and dips occurring once per period. Examining the location of those humps and dips highlighted by the arrows in the RV plot in the bottom right panel reveals that, once again, the greatest uncertainty reductions (e.g., at \gap~=~0.95$\times$\pmax~and \gap~=~1.95$\times$\pmax) were achieved by RVs covering the steepest part of the RV curve, where stellar acceleration is the highest. Due to HD 219077's high eccentricity, only one portion of the orbit experiences this high acceleration, resulting in a single pair of humps and dips in the uncertainty reduction. In contrast, the near circular orbit of HD~134987~c has both approaching and receding portions of the orbit contributing similar acceleration along our line of sight. Thus, data covering either portion produces similar uncertainty reduction patterns, leading to two groups of uncertainty humps and dips per orbital period. This confirms that future observations should ideally be conducted at phases of maximum stellar acceleration to achieve the best orbital period uncertainty reduction.
    
    \subsection{Full Simulations}
    \label{sec:fullsims}
    
    The exploratory simulations were conducted independently for each variable, with the uncertainty analysis performed without considering the relationships among the three variables. Here, we conducted a suite of simulations to capture a comprehensive view of \pmax~uncertainty reduction in relation to all three variables for HD~134987. Each variable's value was varied within a specified range and step size as before, except that for each iteration, two other variables were varied as well. This setup created a 3D grid with each variable represented on one of the axes. We then ran full simulations to extract \pmax~uncertainty for each combination of the three variables at every grid point in the 3D space, resulting in a 4D space with uncertainty values across all grid points.

    We used the original RVs for the HD~134987 system as an example of an archival dataset with which new RVs at future epochs would be created. The procedure of generating the synthetic data was similar to before, but with adjusted step sizes and ranges to better capture relationships among all three variables and their long-term behaviors. To reduce computational cost, we lowered the simulation resolution by increasing the step size for each variable. The 3D grid was formed with the \num~varying from 6 to 200, in steps of 1, \gap~from 0.25$\times$\pmax~to 12$\times$\pmax~in steps of 0.25$\times$\pmax, and \cov~from 0.5$\times$\pmax~to 12$\times$\pmax, in steps of 0.5$\times$\pmax. As in the exploratory and testing runs, we repeated the modeling process five times at each grid point and averaged the results to minimize scatter due to broad MCMC posteriors. In total, 1,123,200 scenarios were run and fit using {\sc RadVel} for the full simulation. 
    
    To model the simulation results, we employed a Random Forest Regressor model to simultaneously analyze the impact of all three variables on \pmax~uncertainty using the Python machine learning framework {\sc Scikit-learn} \citep{pedregosa2011}. The modeling allows us to find the best fit through three dimensions, gain insight into the mutual dependence of the modeled variables, and their relative importance in the model. Random forest is a tree-based ensemble learning method that operates by constructing multiple decision trees during training and aggregating their results to make predictions, either for classification or regression tasks. Each decision tree itself is a single, independent modeling unit built on a random subset of the data and predictor variables (in our case, only three: \num, \gap, and \cov). This randomness helps reduce overfitting, as individual trees may capture noise, but the aggregation of results from all the decision trees smooths out the overall prediction. At each iteration, each decision tree decides on the best available variable to use at the current tree node and splits the subset of data assigned to each tree in a way that minimize the variance within that subset. Such a splitting strategy occurs iteratively at all nodes within a tree for all trees in the model until it meets certain criteria such as the maximum number of tree depth, minimum number of samples left in the each node, minimum reduction in the chosen performance metric such as the mean squared error (MSE) between the current level within the tree to the previous level, and many others. For regression tasks, each tree outputs a prediction by averaging the values in the relevant leaf node (tree nodes after which no more splitting occurs) and the Random Forest model finally takes the average of the predictions from all the individual trees to provide the final output. Since our dataset is relatively large but not too complicated, this model suits our modeling purpose perfectly as it is able to scale well to large datasets, prevent overfitting to the data, and yield accurate predictions with a relatively simple algorithm.

\begin{figure*}[tbp]
    \includegraphics[width=\textwidth]{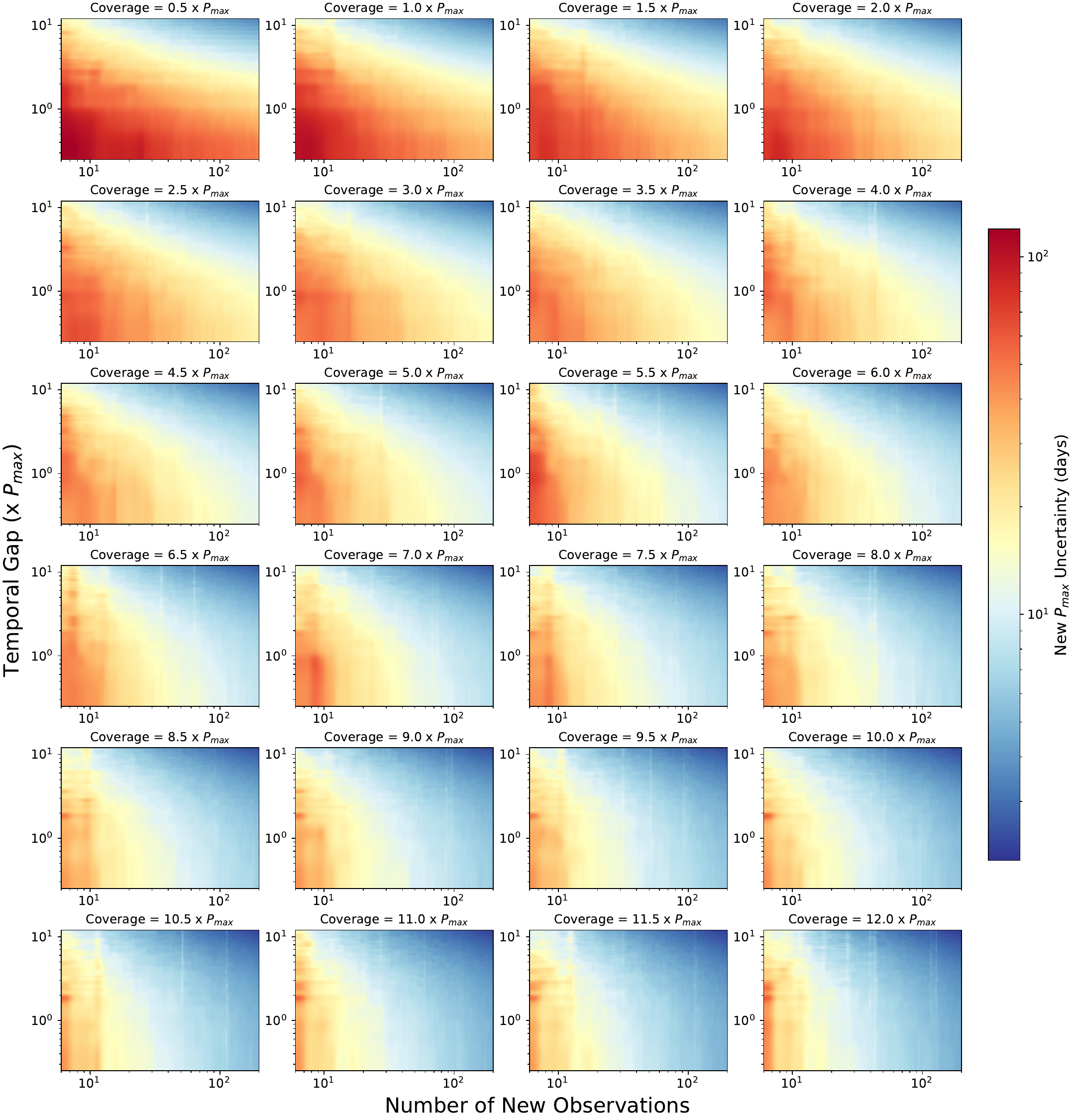}
    \caption{The best-fit random forest model of the full simulation data for HD 134987 c, with the \num~variable on the horizontal axis, the \gap~variable on the vertical axis, and the \cov~variable represented across the panels. The \gap~and the \cov~variables are expressed in unit of \pmax, the orbital period of the outermost planet. Derived \pmax~uncertainties are color coded in each panel, with higher and lower ends of the uncertainty range coded by red and blue, respectively. Fluctuations within the contour are due to the final resolution on both axes and the nature of the random forest model that is completely data informed and therefore produces a relatively rough surface. The \pmax~uncertainty before the simulations is 410 days.}
    \label{fig:rfmodel1}
    \end{figure*}

    To train the model, all samples in our dataset were reshuffled and divided into 75\% training set and 25\% testing set. The model was then trained on the training data using 10-fold cross-validation while performing a grid search for hyperparameter tuning and MSE as the criterion to measure the quality of the split at each node. The hyperparameters tuned along with the values selected for the best model, are included in Table~\ref{tab:hyperparam}. The best-fit model yields a root mean squared error of 2.08 and 1.85 days on the training and testing data, respectively. The training and testing data errors yield similar low values which suggests the model has been well-trained. Once the model was trained, we applied the entire dataset to the best-fit model, with results smoothed by our model shown in Figure~\ref{fig:rfmodel1}. Since our data is in four dimensions, we plot the \num~variable on the horizontal axis and the \gap~variable on the vertical axis, with the \cov~variable represented in each panel, and the derived \pmax~uncertainty color coded in each panel. Once again, the three variables are expressed in units of the orbital period of the outermost planet, \pmax. Due to the entirely data-driven nature of the Random Forest algorithm, the final smoothed model of the simulation data still exhibits some localized fluctuation due to noise and therefore does not produce a completely smoothed contour as a traditional parametric model would. However, the model yields a relatively smoothed contour without outliers, showing the general relationship among the three variables. We additionally include two plots in Appendix~\ref{appx:figures}, with Figure~\ref{fig:rfmodel2} and Figure~\ref{fig:rfmodel3} showing the \gap~variable and the \num~variable in the panels, respectively.
    
    \begin{deluxetable}{lr}[tbp]
    \tablecaption{Values Used For Hyperparameter Tuning.
    \label{tab:hyperparam}}
    \tablewidth{\columnwidth}
    \tablehead{
        \colhead{Hyperparameters} &
        \colhead{Values}}
    \startdata
    max depth & 2, 4, 6, 8, 10, \textbf{20}, 30, None \\ 
    min samples leaf & 1, \textbf{2}, 5, 10, 15, 20  \\ 
    min samples split & 2, 4, 6, 8, \textbf{10}, 12, 15, 20 \\
    max features & \textbf{1}, 2, 3 \\
    n estimators & 100, 200, 300, \textbf{400}, 500, 600, 700 \\
    min impurity decrease & \textbf{0}, 0.01 \\\hline
    \enddata
    \tablecomments{Hyperparameter values selected for the best-fit model are highlighted in bold. Details of each hyperparameter see {\sc scikit-learn} documentation page.}
    \end{deluxetable}
    
    Given that the simulation resolution for the \gap~variable is much lower compared to that in the exploratory runs, the effects of high and low \pmax~uncertainties for the \gap~variable, which indicate different starting timestamps of simulated runs, are not as clearly visible as before. In addition, the \pmax~uncertainty fluctuation is greatest when the \num~variable is at a low value and becomes smoother as \num~increases, as can be seen in all three figures. This makes sense since a smaller number of future observations would not be able to constrain the orbit as well, and the results from the MCMC process would be highly variable compared to cases with large amount of RVs at future epochs. In all three figures, the overall \pmax~uncertainty decreases monotonically with increasing values of all three variables. Specifically, the rate of uncertainty reduction happens fastest when the \num, \gap, and \cov~variables are all small, and then gradually slows to an almost static state when only two of the variables have large values. This is evident in all three figures, as there is little change in the color and distribution of the contour in the bottom half of the panels. This suggests that there is a stopping threshold for the three variables when planning observations. Once the threshold is crossed, further delaying the start of an observational campaign, increasing the baseline coverage, or greatly increasing the number of observations or cadence, would not yield a similar amount of \pmax~uncertainty reduction as before, and the value in waiting longer and utilizing more observing resources would reach a point of diminishing returns. Such a threshold can be identified at the ``elbow" point of the uncertainty reduction curve in a 2D plot, or the ``cliff" in a 3D contour plot, beyond which the gradient of uncertainty drastically reduces.
    
\begin{figure}[tbp]
        \includegraphics[trim=0 70 20 100,clip,width=1.0\columnwidth]{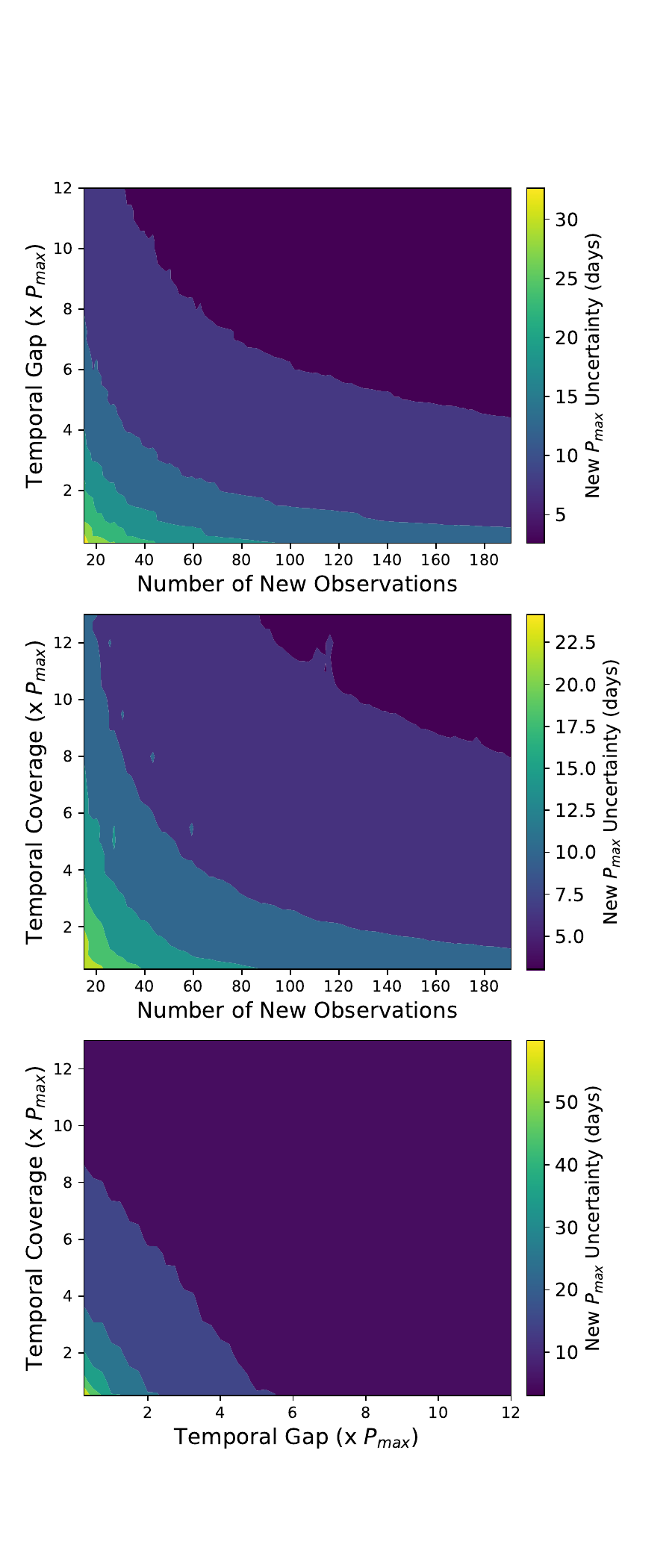}
        \caption{2D Partial Dependence Plots (PDP) of the three variables from the full simulation. A PDP with two variables shows the mutual variable interaction and effects on the model output while marginalizing other variables. Top panel: \gap~vs \num; middle panel: \cov~vs \num; bottom panel: \cov~vs \gap. Darker color represents lower \pmax~uncertainties values. The \pmax~uncertainty before the simulations is 410 days.}
    \label{fig:pdp}
    \end{figure}

    The location of this stopping threshold can be seen more clearly with the help of partial dependence plots (PDP). A PDP shows the dependence between the model output, in this case \pmax~uncertainty, and the input variable of interest, marginalizing over the values of all other variables. If two variables are selected for a PDP, then the interaction between the two variables and their effects on the model output can be revealed with a 2D contour. Here in Figure \ref{fig:pdp}, we show 2D PDPs between the three variables from the full simulation. The top, middle, and bottom panels show the PDPs for \num~and \gap, \num~and \cov, and \gap~and \cov, respectively. Evidently, \pmax~uncertainty reduces towards the upper right corner as values for both variables increase, but the rate decreases with the magnitude of both variables. In the top and the middle panels, \pmax~experiences a larger range of uncertainty reduction at lower \gap~and \cov~values as \num~increases, and the reduction curve becomes flatter as \gap~and \cov~increase towards high values, making observing scenarios with high \gap~or \cov~less rewarding. For example, at \gap~$= 4\times$\pmax in the top panel, it is apparent that observing with 200 observations is no better than observing with only 80 observations. At \gap~$= 5\times$\pmax, 40 observations should be sufficient. Similar conclusions can be drawn for the \cov~variable in the middle panel: for the \cov~$= 2\times$\pmax scenario, there is no need to carry out large number of observations, as 60 is enough. In the bottom panel, where the interaction between the \gap~and the \cov~variables are displayed, the \pmax~uncertainty reduction behaves similarly, except in a more linear fashion in this parameter space. The stopping threshold, where further increasing the variable yield little additional value, is more obvious: beyond \gap~$= 6\times$\pmax, for example, increasing the temporal gap gains little, and the \pmax uncertainty becomes insensitive to \cov~variables at all.

    In this particular case, the \pmax~uncertainty reduction does not level off until the \gap~and \cov~variables reach large values. Given the long period of the HD 134987 c planet ($\sim$5000 days), the concern for diminishing returns for these two variables is practically irrelevant, as it is highly unlikely that the next RV monitoring campaign would start nearly a century later or last for a similar duration. However, this might not hold true for other targets with shorter periods. It is important to highlight that the threshold for reaching the uncertainty plateau for the \num~variable is a key factor to consider when planning observations, as one should aim for the most efficient observing strategy and avoid oversubscribing valuable time on precision RV instruments unnecessarily.
    
\begin{figure}[tbp]
    \includegraphics[trim=20 0 30 30,clip,width=\columnwidth]{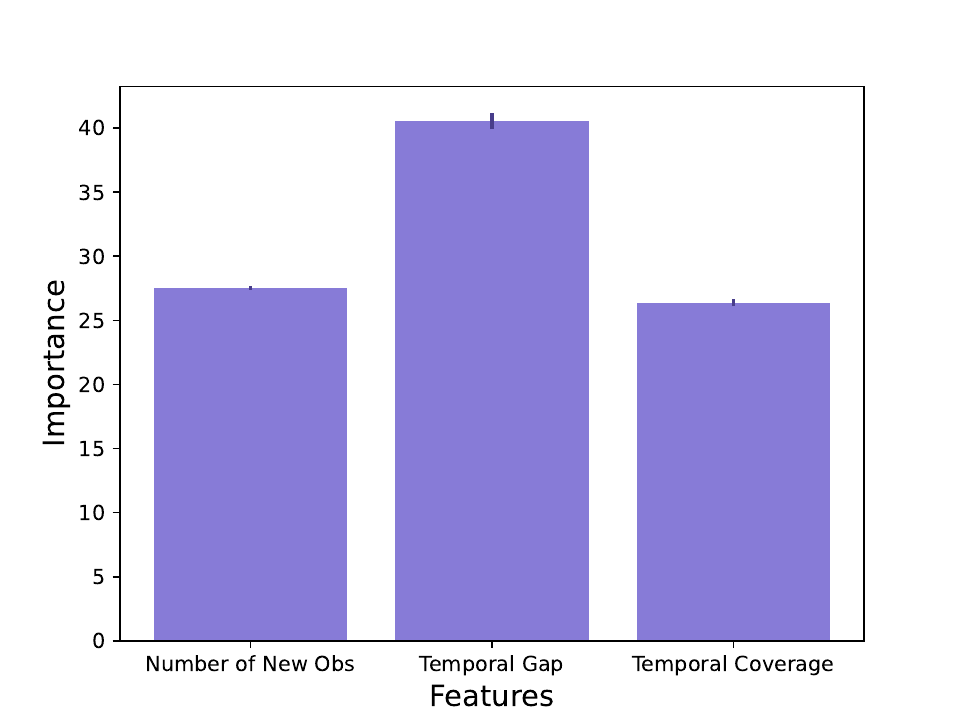}
    \caption{Feature importance of the three variables, \num, \gap, and \cov~variables in the best-fit random forest model on the full simulation data. Each bar represents the mean importance from all permutations carried out, with the vertical line at the top of each bar showing the standard deviation of the importance for that variable/feature. The higher the bar is, the more importance that variable is to our model.}
    \label{fig:featureimportance}
    \end{figure}

    Another way to examine each variable's impact on our model is to look at the importance of each variable. Here, we calculate the feature importance for \num, \gap, and \cov~within the selected best-fit model using permutation feature importance. This model inspection technique measures the importance of each variable to the overall model performance by randomly shuffling data of the variables under inspection to break relationships between target and variables. If the permutation result shows a large decrease in model performance for a particular variable compared to a baseline value, then it suggests the model is highly dependent on that variable and the variable has high importance within the model. The higher the feature importance value of a variable, the greater its impact within the model. This technique is more robust than other methods, such as impurity-based feature importance, as its calculation is model-agnostic and not biased toward features with high cardinality. The feature importance results of our best-fit model are shown in Figure~\ref{fig:featureimportance}. Each bar represents one of the variables, with the standard deviation of each importance value indicated by a vertical line at the top of each bar. The figure suggests that all three variables contribute meaningfully to the model, given the non-negligible feature importance values, which is reassuring. Furthermore, it appears that the \gap~variable has the greatest importance, while the \num~ and \cov~variables have comparable importance. Because of this, the \gap~variable possesses the highest predictive power within our model and has the highest influence on the model performance, followed by \num~and \cov~variable. For instance, the model indicates that, on average, holding \num~and \cov~constant while varying \gap~leads to a larger reduction in \pmax~uncertainty than other scenarios where \cov~is varied while holding \num~or \gap~constant, or \num~is varied with \gap~and \cov~fixed. This finding indicates that, all else being equal, delaying future observations to increase the temporal gap between the last observation and the upcoming campaign is relatively more effective in reducing \pmax~uncertainties. 

    
\section{Alternative Simulations}
\label{sec:app}

Of course, knowing the optimal observing strategy and the corresponding expected \pmax~uncertainty reduction performance before an observing campaign is of particular importance to observers. The full simulation in earlier sections provides valuable insights into \pmax~uncertainty behavior concerning all three variables simultaneously, as well as the relative importance of each variable. However, such a full simulation is computationally expensive and typically impractical to conduct for all targets before the campaign begins. In addition, the full simulation does not take into account realistic constraints, such as future mission start times before which observations have to be taken. In this section, we therefore offer examples of shorter simulations with realistic constraints on the start and end times of future observations in the context of upcoming space-based direct imaging missions, such as the Roman telescope.
    
    \subsection{Simulation Objects}
    \label{sec:simobject}

    For the short simulations, we selected three targets that are currently high-priority candidates for the Roman CGI\footnote{The complete high-priority target list for Roman CGI can be found at the Imaging Mission Database: \url{https://plandb.sioslab.com/index.php}.}: HD 134987, the same target as before \citep{jones2010}; 47 UMa, a 3-planet system popular in the direct imaging community \citep{butler1996a,fischer2002,gregory2010}; and 14 Her, a 2-planet system with one very long period planet \citep{butler2003, wittenmyer2007, bardalez2021, rosenthal2021}. Similar to the previous simulations, estimates of orbital parameters are required to create synthetic RVs. Instead of using parameters from a single prior publication, we compiled all publicly available RVs from various sources and modeled them together to derive the most up-to-date orbital parameters for all three targets using {\sc RadVel}. The most current orbital parameters for the HD 134987, 47 UMa, and 14 Her systems are provided in Table \ref{tab:param}.

    \begin{deluxetable*}{ccccccccc}[tbp]
    \tablecaption{Most Up-to-date System Parameters of HD 134987, 47UMa, and 14 Her.
    \label{tab:param}}
    \tablehead{
        \colhead{System} & 
        \colhead{Planet} &
        \colhead{$P$ (days)} &
        \colhead{$e$} &
        \colhead{$\omega$ (deg)} &
        \colhead{$T_{p}$ (BJD)} &
        \colhead{$K$ (m~s$^{-1}$)} &
        \colhead{$a$ (au)} &
        \colhead{$M_{p}\mathrm{sin}i$ ($M_{\rm J}$)}
    }
    \startdata
    \multirow{2}{*}{\shortstack{HD 134987$^{a}$}} 
     & b & $258.25 \pm 0.025$ & $0.2318 \pm 0.0051$ & $355.2 \pm 1.4$ & $2,459,110.96 \pm 0.96$ & $50.03 \pm 0.29$ & $0.818 \pm 0.012$ & $1.62^{+0.047}_{-0.048}$ \\
     & c & $6348^{+71}_{-53}$ & $0.099^{+0.031}_{-0.032}$ & $291^{+17}_{-14}$ & $2,458,512^{+420}_{-240}$ & $11.03 \pm 0.29$ & $6.92 \pm 0.11$ & $1.061 \pm 0.042$\\
     \hline
     \multirow{3}{*}{\shortstack{47 UMa$^{b}$}} 
     & b & $1075.61^{+0.79}_{-0.67}$ & $0.0312^{+0.0071}_{-0.0073}$ & $335^{+15}_{-14}$ & $2,458,379^{+46}_{-44}$ & $47.31^{+0.50}_{-0.46}$ & $2.06^{+0.032}_{-0.033}$ & $2.395 \pm 0.079$ \\
     & c & $2290 \pm 11$ & $0.255^{+0.057}_{-0.070}$ & $86 \pm 11$ & $2,457,882^{+64}_{-58}$ & $7.61 \pm 0.45$ & $3.407^{+0.054}_{-0.055}$ & $0.478^{+0.031}_{-0.030}$\\
     & d & $16288^{+950}_{-340}$ & $0.376^{+0.075}_{-0.081}$ & $66^{+13}_{-12}$ & $2,451,397^{+280}_{-370}$ & $11.87^{+1.00}_{-0.85}$ & $12.64^{+0.49}_{-0.34}$ & $1.38^{+0.16}_{-0.13}$ \\
     \hline
     \multirow{2}{*}{\shortstack{14 Her$^{c}$}} 
     & b & $1766.71^{+0.58}_{-0.59}$ & $0.3652^{+0.0035}_{-0.0034}$ & $22.60^{+0.57}_{-0.56}$ & $2,456,673.4^{+2.3}_{-2.2}$ & $89.47 \pm 0.42$ & $2.835^{+0.040}_{-0.041}$ & $4.86 \pm 0.14$ \\
     & c & $19751^{+4700}_{-590}$ & $0.372^{+0.058}_{-0.026}$ & $21.3^{+3.3}_{-1.5}$ & $2,451,427^{+160}_{-140}$ & $43.8^{+2.7}_{-1.8}$ & $14.26^{+2.1}_{-0.52}$ & $5.34^{+0.53}_{-0.31}$\\
    \enddata
    \tablecomments{The $\omega$ values in each row of the table are those of the star, not of the planets. Stellar mass for each system was taken from \citet{rosenthal2021}.}
    \tablenotetext{a}{Stellar mass $M_*$ = $1.093 \pm 0.047$ $M_{\sun}$. RVs included: HARPS \citep[pre- and post-upgrade;][]{trifonov2020}, HIRES \citep[pre- and post-upgrade;][]{rosenthal2021}, and UCLES \citep{jones2010}. Current best estimate for planet c's period from NASA Exoplanet Archive is $5960^{+170}_{-150}$ days from \citet{rosenthal2021}. A separate work by \citet{wittenmyer2020b} reports a period of $5358 \pm 31$ days for planet c.}
    \tablenotetext{b}{Stellar mass $M_*$ = $1.005 \pm 0.047$ $M_{\sun}$. RVs included: APF-Levy \citep{rosenthal2021}, ELODIE \citep{naef2004}, HIRES \citep[post-upgrade;][]{rosenthal2021}, HJS-Coude \citep{wittenmyer2009}, HET-HRS \citep{wittenmyer2009}, and Lick-Hamilton \citep{rosenthal2021}. Current best estimate for planet d's period from NASA Exoplanet Archive is $14002^{+4018}_{-5095}$ days from \citet{gregory2010}.}
    \tablenotetext{c}{Stellar mass $M_*$ = $0.969 \pm 0.042$ $M_{\sun}$. RVs included: APF-Levy \citep{rosenthal2021}, ELODIE \citep{naef2004}, HIRES \citep[pre- and post-upgrade;][]{rosenthal2021}, HJS-Coude \citep{wittenmyer2009}. Current best estimate for planet c's period from NASA Exoplanet Archive is $15732^{+2654}_{-1896}$ days from \citet{feng2022}.}
    \end{deluxetable*}

    \subsection{Simulations and Result}
    \label{sec:simresult}

    Although 47 UMa has three planets of high interest to the direct imaging community, we focus only on the outer most planet, d, for the observing strategy simulation for \pmax~reduction, and apply the same focus to the other two systems as well. The synthetic RV generation follows the same method as before, but the simulation setup is slightly modified. Instead of creating a full grid of simulations based on the three variables \num, \gap, and \cov~to test uncertainty changes without enforcing a fixed starting and ending time for the observing campaign at future epochs, these shorter simulations were all set to end at the same future date. This design mimics the need to complete precursor observations before future mission begins. The primary difference between these short simulations is their varying starting times. For each target, we conducted eight different runs under the scenario of precursor RV observations for Roman targets, with all runs ending around February 1st, 2028, assuming a launch date of May 2027 and accounting for the engineering time before the Roman's science observations commence. The first run begins on February 1st, 2024, with subsequent runs starting 6 months after the previous run's start date. All other simulation setup details remained consistent with prior simulations. Because of the fixed ending time, each successive run has a larger \gap~and smaller \cov~value than the previous runs. However, given the short observing baseline and large \pmax~values in this case, these differences are quite small. Details of the eight runs for the three targets are summarized in Table~\ref{tab:rundetails} in Appendix~\ref{appx:tables}. Once the simulations were complete, we compiled the results from all the runs together for each target and created a best-fit smoothed model around each run's data along with a confidence interval, similar to what was done in the left panels of Figure~\ref{fig:case2}. To fully capture the \pmax~uncertainty behavior, including the initial rapid decline and other nonlinear features, we used the Generalized Additive Model (GAM) to model the short simulation results. 

\begin{figure*}[tbp]
        \begin{center}
            \begin{tabular}{cc}
                \includegraphics[trim=0 0 0 0,clip,width=0.48\textwidth]{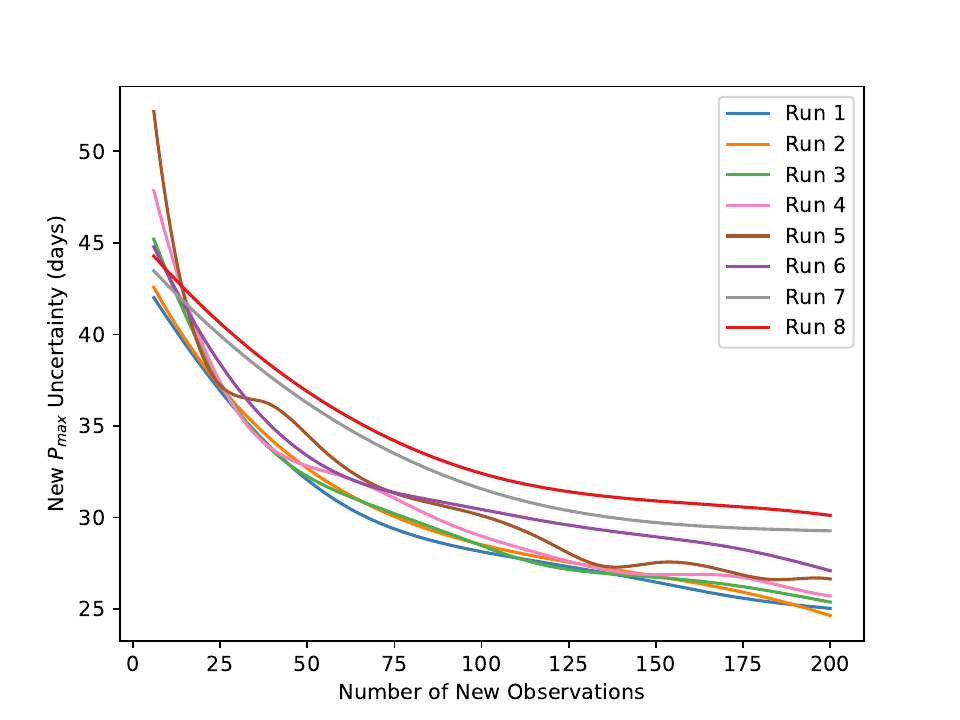} &
                \includegraphics[trim=0 0 0 0,clip,width=0.48\textwidth]{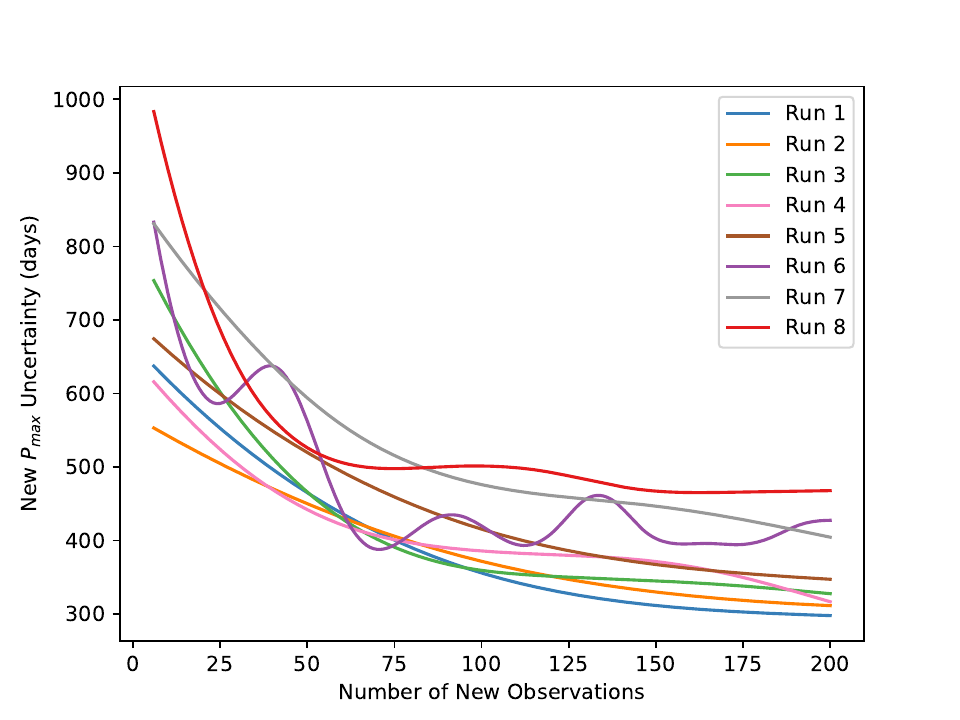} \\
                \includegraphics[trim=0 0 0 0,clip,width=0.48\textwidth]{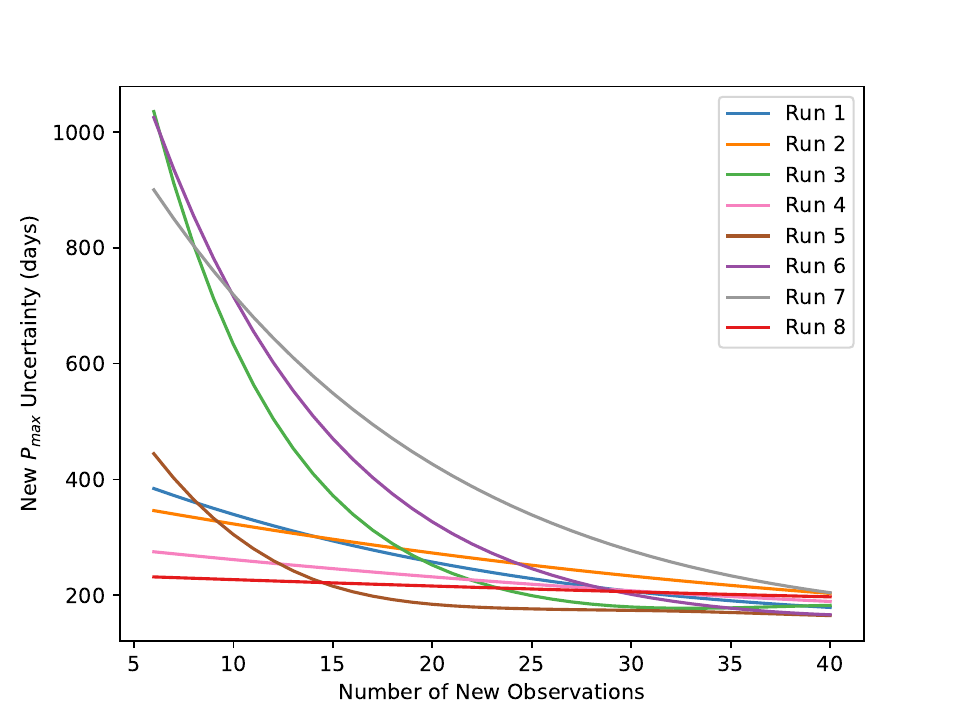} &
                \includegraphics[trim=0 0 0 0,clip,width=0.48\textwidth]{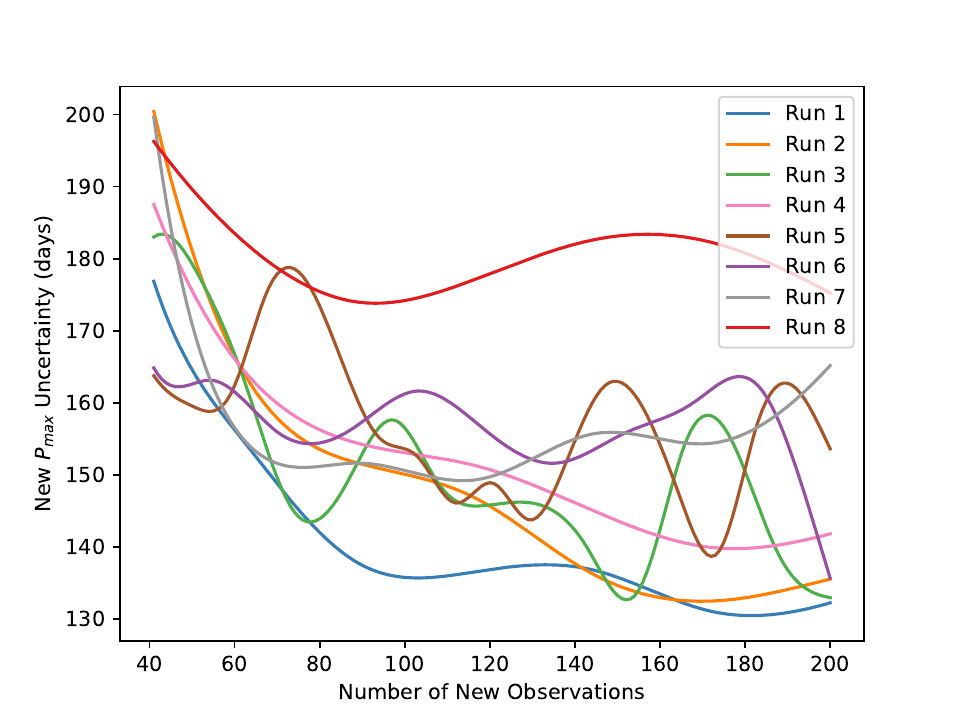}
            \end{tabular}
        \end{center}
        \caption{Short simulation results for HD~134987 (top left), 14 Her (top right), and 47 UMa (bottom panels) using GAM. All eight runs for each target are included with confidence intervals omitted to avoid figures being too cluttered. Results for 47 UMa was split into left (\num$\le$40) and right (\num$>$40) panels to reveal differences between runs at larger \num~values. The \pmax~uncertainties before the simulations for HD 134987, 14 Her, and 47 UMa are 62, 2600, and 640 days, respectively.}
    \label{fig:gam}
    \end{figure*}

    GAM is a semi-parametric model based on the Generalized Linear Model, which models the relationship between the target, or dependent variable $Y$ and one or more predictor, or independent variables $x_{n}$ by allowing for nonlinear relationships. GAM takes the form of:

    \begin{equation}
        \label{eqn:gam}
         g(E(Y)) = \beta_{0} + f_{1}(x_{1}) + f_{2}(x_{2}) + ... + f_{n}(x_{n})
    \end{equation}
    where $E(Y)$ is the expected value of $Y$, $g()$ is the link function relating the expected value of the target variable to predictor variables, $\beta_{0}$ is the intercept term; and $f_{n}()$ represents feature functions of the predictor variables, which can take on parametric or non-parametric forms. In this case, the feature functions used here are penalized B splines that would allow us to model nonlinear relationships without specifying a particular functional form. GAM additively combines the effects of individual smoothing functions and includes regularization to help prevent overfitting. This flexibility and ability to model highly nonlinear relationship makes GAM an excellent tool for data with unknown degree of nonlinearity. Here, we used the {\sc PyGAM} package \citep{serven2018} to model the short simulation results. For the spline terms, a penalty on the second derivative was applied to regularize the model and produce smoother results. The target variable $Y$ must be specified from a family of exponential distributions; in our case, a gamma distribution with a log transformation link function was chosen. A grid search was performed to find the best penalizing values as well as the optimal number of splines for the spline terms using generalized cross validation (GCV). The smoothing penalty was varied across 30 values spaced logarithmically between -2 and 2, with larger values imposing stronger penalties. The number of splines was varied between 5 and 30 in steps of 5, with higher numbers able to fit more complex structures but also increases the risk of overfitting. The grid search selects the model with the lowest GCV score as the best-fit model for each run. We present the modeling results for all eight runs of all three targets in Figure~\ref{fig:gam}. 

    In all runs for the three targets, the \pmax~uncertainty measurement improves as expected. However, the rate of reduction is less drastic compared to the exploratory or full simulations due to the much shorter temporal baseline. The \pmax~uncertainty behaviors differ across the targets due to variations in the original datasets (e.g., total previous baseline, number of previous observations) and the unique characteristics of each planet. For instance, the uncertainty reduction effect for HD 134987 c is not as significant as for 47 UMa d, or 14 Her c, since HD 134987 c has a relatively shorter orbit and therefore its orbital uncertainty is already fairly well constrained. Several key details can be seen from the figures. First, as in the exploratory and full simulations, the uncertainty curve for each run flattens at larger \num~values, indicating that additional observations may not necessarily add much more values in constraining the period. For example, in all cases, there is little gain in reducing \pmax~uncertainty once the number of new observations exceeds 100. Second, due to the data-driven nature of the spline terms, fluctuations in data will be reflected in the model. Such variations, especially at higher \num~values as seen in Run 6 of 14 Her (in the upper right panel of Figure~\ref{fig:gam}) and runs of 47 UMa (in the lower right panel of Figure~\ref{fig:gam}), do not come from sampling along different parts of the phase curve as we saw in Figure~\ref{fig:case2}, but rather arise from the relatively noisier MCMC solutions. This is likely due to the complexity of the planetary system. For example, the three-planet 47 UMa system exhibited higher level fluctuations than the two-planet 14 Her system. This suggests model unpredictability at larger \num~values for more complex systems, where \pmax~uncertainty has already reached the performance plateau, further suggesting limited value in additional observations in this case. Finally, for targets with results that do not exhibit high level of uncertainty fluctuations, it can be seen that runs with earlier start dates tend to outperform those with later start dates. Runs that began their observations sooner, with smaller \gap~values, generally show better performance than those that started later. This trend holds even for the noisier case of 47 UMa, though the performance margin between the best and the worst runs vary by target and dataset characteristics. This result suggests that, for shorter simulations with a fixed end time, beginning observations as early as possible improves performance by increasing the temporal coverage of future data and minimizing the temporal gap between old and new data. It is worth noting that this conclusion does not contradict results from the full simulations, where the \gap~variable was the most influential factor under the assumption that adjusting the \gap~ or \cov~variables does not reduce the total temporal baseline (\gap~plus \cov). However, we do not have such an assumption here for the short simulations due to the constraint of having a fixed observing end time. A later start date would result in a shorter total temporal baseline, resulting in poorer performance.

    \subsection{An RV Data and Orbits Repository for Direct Imaging}
    \label{sec:repo}

    To address the growing need for the most refined orbital parameters of direct imaging candidates from RV observations for purposes such as mission planning and RV observing simulations presented in this paper, we are establishing a \href{https://github.com/sblunt/roman-orbits}{Github repository}. Our goal is to host all publicly available RV data for relevant direct imaging candidates, along with fitted RV models, providing the community with the most up-to-date orbital parameters estimates. Many publications model and publish only the RVs observed within their specific observing programs, resulting in partial updates on a target's orbital parameters. This practice often misses the opportunity to provide the most refined parameters by not incorporating all previously available data. Our repository aims to offer the most comprehensive RV models of direct imaging targets with all publicly available data aggregated in one place. We plan to automate the modeling process using {\sc RadVel}, employing a shell script to organize and concatenate all input data before feeding it into the modeling setup file. This will expedite follow-up modeling when new data are added to the repository and ensure that all data are modeled in a consistent manner. The repository is currently in its early stages, and collecting all data and performing RV modeling on all the targets will take time. Currently, we plan to include all targets on the Roman CGI candidate list and, if resources allow, incorporate astrometry data and expand the repository to include targets for missions such as the Habitable Worlds Observatory, or even non-imaging missions. We welcome community members interested in joining and contributing to this effort. All data and the latest RV models for the three targets discussed in Section~\ref{sec:simobject} are already included in the repository.


\section{Discussion}
\label{sec:discuss}

We performed simulations to explore the orbital period uncertainty behavior of the outermost planet in a system, focusing on the influence of three RV observational variables: the number of future observations, the temporal gap between the old and the new data, and the temporal coverage of the new observations. Synthetic RVs were generated to mimic real-world data, including observational white noise and uneven temporal sampling. However, some key observational constraints were not included, and we briefly discuss their potential effects on our simulations here.

Temporally correlated red noise can significantly contribute to the total error budget for RV observations if the stars in question exhibits strong intrinsic stellar variability. Mitigating correlated noise is an active area of research within the RV community, particularly for the EPRV observations aimed at detecting Earth-like planets with small RV signals \citep{dumusque2011,fischer2016,hall2018,chaplin2019,crass2021,luhn2023,gupta2024}. However, for our simulations, which focus on long-period, high-amplitude gas giant planets orbiting mature, non-active stars, short-term low-amplitude stellar activity signals are less relevant here, and the treatment of correlated noise is generally unnecessary. We refer readers to EPRV studies for more on the impact of correlated noise on RV observations. We expect that the exclusion of red noise in our simulated data may slightly underestimate the derived errors of the orbital parameters and therefore the \pmax~uncertainty curves or contours would be slightly inflated or have a less dramatic initial decline. However, this would not affect the overall results.

Seasonality is another common issue for observations, as targets may be unobservable in the night sky for part of the year. The loss of observing opportunities due to target visibility will effect uncertainty outcomes. In our simulations, we assumed a simplified approach where observations are not impacted by target visibility. This assumption was made to focus on the effects of the three main variables \num, \gap, and \cov, without introducing another confounding variable that could overcomplicate the model. We expect that excluding seasonality may slightly underestimate derived parameter errors. This is especially true if the target has an orbit similar to the high eccentricity case of Figure \ref{fig:case2}, in which case, missing the optimal observing time due to seasonal visibility may have a significant impact on the uncertainty reduction. In these cases, we recommend conducting observations at the next best time, such as when \gap~is around 0.90$\times$\pmax~of the HD 219077 case.

Disruption to observing programs due to weather, forest fire, instrument failures, or interruptions from other observing programs could negatively impact an observing campaign. Since our simulations did not assume a specific observing facility, we did not account for annual average weather conditions at any particular observatory. The randomness we introduced into the temporal sampling of our synthetic RV data partially mimics the weather unpredictability, though its effect may vary based on actual conditions at a given location. We do not expect this to have a noticeable impact on our simulation results.


\section{Conclusions}
\label{sec:conclude}

Long-period gas giant planets that are potential candidates for future direct imaging missions, such as Roman, typically have poorly constrained orbital parameters due to the long observational baseline required to fully characterize their orbits. The large uncertainties associated with these parameters will have an impact on the quality of future imaging missions, which will depend on the available orbital information extracted from existing RV data. Given the limited time available on precision RV facilities, there is therefore an urgent need for the RV community to plan an efficient precursor observing campaign to provide the best constraints on orbital parameters of candidates before the start of these imaging missions. In this work, we explored the effects of the \num, \gap, and \cov~variables on the behavior of \pmax~uncertainty through a suite of full simulations and provided optimal observing solutions under realistic constraints of these three variables through a set of short simulations. We summarize our findings below:

\begin{enumerate}
    \item For the purpose of refining orbits of cold gas giant planets that are candidates of future direct imaging missions, the use of EPRV facilities is unnecessary and the use of RV facilities with precision on the order of meter-per-second suffices.
    \item For future observations with partial temporal coverages (\cov$<$1$\times$\pmax), the start time of an observing campaign, determined by the \gap~variable, is especially important since the RV phases covered by observations influence the extent of \pmax~uncertainty reduction. Specifically, future RVs that cover the phase where the RV curve's gradient is steepest provide better \pmax~uncertainty reduction compared to observations along other RV phases.
    \item All three variables have significant contributions to \pmax~uncertainty reduction, with the \gap~and \cov~variables having the most and least predictive power, respectively, on the model's performance. On average, the \gap~variable has the largest influence on the \pmax~uncertainty model performance, and fixing \num~and \cov~while varying \gap~tends to result in the greatest reduction in \pmax~uncertainty. With all else being equal, it is preferable to delay the start of future observations, allowing for an increased temporal gap between the previous and upcoming campaigns, which could deliver better uncertainty reduction results, given that there is no strict end time for future observations. This would be the case for low priority stars that are not part of the target lists of any future missions. Removing them from the observing queue and saving time for higher priority targets would not necessarily harm their orbit refinement progress.
    \item If the future observing campaign is constrained by the strict program end time, it can achieve a better \pmax uncertainty reduction by beginning observations as soon as possible, rather than delaying as in cases without a hard end time. Within the given observing window, increasing the temporal coverage of future observations as much as possible enhances \pmax~uncertainty reduction performance. This would be the case for precursor observations for future direct imaging missions where orbit refinement needs to be performed before mission target lists can be finalized. Therefore, such observations should be assigned with high priorities and commenced as early as possible. In this case, when the total temporal baseline (including all old and new data) was fixed, smaller \gap~and therefore larger \cov~on average contributed to better \pmax~uncertainty performance. This highlights \gap~and \cov~being crucial variables, and how \gap~can affect model performance differently under different conditions.
    \item Obtaining too many observations within a limited time frame, delaying future observations for too long, or conducting an observing campaign for too long is unnecessary. The \pmax~uncertainty eventually plateaus past a certain stopping threshold, where further increases in these variables do not significantly improve uncertainty reduction. One should look for the ``elbow" point of the uncertainty curve beyond which uncertainty decrease starts to level off. Careful observational planning is necessary to avoid oversubscribing valuable RV facilities for observations that offer limited gains in value.
    \item We are developing a Github repository\footnote{The Github repository for storing all RV data and modeling of imaging targets can be accessed here: \url{https://github.com/sblunt/roman-orbits}.} aimed at housing all publicly available RV data and their latest models to provide the best orbital parameter estimates for direct imaging candidates. Currently, priority is given to Roman targets, with plans to expand the repository to include other high-priority targets for various missions. The code for the simulations presented in this paper are available as well\footnote{The Github repository for code used to run the simulations can be accessed here: \url{https://github.com/zhexingli/RVRefine}.}.
\end{enumerate}

Refinement of orbital period uncertainties for long-period planets within known planetary systems that will become direct imaging candidates is a critical task that cannot be overlooked. The orbital information provided by RV data not only affects the precision of predicted planetary locations at future epochs, but also influences the scientific yield of DI, which heavily depends on RV measurement precision. A good understanding of how future RV observations could enhance orbital parameter measurement precision, and to what extent each underlying variable impacts will be crucial to the planning of efficient ground-based RV follow-up campaigns without oversubscribing the valuable telescope time. Designing target follow-up campaigns, collecting data, modeling, and maintaining results in a consistent way to provide the most refined orbital parameter estimates requires a community effort and should begin as early as possible to maximize the science return of future DI missions.


\section*{Acknowledgments}

We would like to thank the anonymous referee for providing a swift, detailed, and valuable feedback that greatly improved the quality of this paper. 
Refining orbital parameters for imaging candidates requires substantial effort, and we welcome community contribution to our Github repository. This research has made use the NASA Exoplanet Archive, which is operated  by the California Institute of Technology, under contract with the National Aeronautics and Space Administration under the Exoplanet Exploration Program. C.K.H.\ acknowledges support from the National Science Foundation Graduate Research Fellowship Program under Grant No. DGE 2146752.


\software{{\sc Matplotlib} \citep{hunter2007}, {\sc Numpy} \citep{harris2020}, {\sc pandas} \citep{wes2010}, 
            {\sc pyGAM} \citep{serven2018}, {\sc RadVel} \citep{fulton2018a}, 
            {\sc scikit-learn} \citep{pedregosa2011}, {\sc Scipy \citep{virtanen2020}}}



\begin{thebibliography}{}
\expandafter\ifx\csname natexlab\endcsname\relax\def\natexlab#1{#1}\fi
\providecommand{\url}[1]{\href{#1}{#1}}
\providecommand{\dodoi}[1]{doi:~\href{http://doi.org/#1}{\nolinkurl{#1}}}
\providecommand{\doeprint}[1]{\href{http://ascl.net/#1}{\nolinkurl{http://ascl.net/#1}}}
\providecommand{\doarXiv}[1]{\href{https://arxiv.org/abs/#1}{\nolinkurl{https://arxiv.org/abs/#1}}}

\bibitem[{{Akeson} {et~al.}(2019){Akeson}, {Armus}, {Bachelet}, {Bailey}, {Bartusek}, {Bellini}, {Benford}, {Bennett}, {Bhattacharya}, {Bohlin}, {Boyer}, {Bozza}, {Bryden}, {Calchi Novati}, {Carpenter}, {Casertano}, {Choi}, {Content}, {Dayal}, {Dressler}, {Dor{\'e}}, {Fall}, {Fan}, {Fang}, {Filippenko}, {Finkelstein}, {Foley}, {Furlanetto}, {Kalirai}, {Gaudi}, {Gilbert}, {Girard}, {Grady}, {Greene}, {Guhathakurta}, {Heinrich}, {Hemmati}, {Hendel}, {Henderson}, {Henning}, {Hirata}, {Ho}, {Huff}, {Hutter}, {Jansen}, {Jha}, {Johnson}, {Jones}, {Kasdin}, {Kelly}, {Kirshner}, {Koekemoer}, {Kruk}, {Lewis}, {Macintosh}, {Madau}, {Malhotra}, {Mandel}, {Massara}, {Masters}, {McEnery}, {McQuinn}, {Melchior}, {Melton}, {Mennesson}, {Peeples}, {Penny}, {Perlmutter}, {Pisani}, {Plazas}, {Poleski}, {Postman}, {Ranc}, {Rauscher}, {Rest}, {Roberge}, {Robertson}, {Rodney}, {Rhoads}, {Rhodes}, {Ryan}, {Sahu}, {Sand}, {Scolnic}, {Seth}, {Shvartzvald}, {Siellez}, {Smith}, {Spergel}, {Stassun}, {Street}, {Strolger}, {Szalay},
  {Trauger}, {Troxel}, {Turnbull}, {van der Marel}, {von der Linden}, {Wang}, {Weinberg}, {Williams}, {Windhorst}, {Wollack}, {Wu}, {Yee}, \& {Zimmerman}}]{akeson2019}
{Akeson}, R., {Armus}, L., {Bachelet}, E., {et~al.} 2019, arXiv e-prints, arXiv:1902.05569, \dodoi{10.48550/arXiv.1902.05569}

\bibitem[{{Bailey} {et~al.}(2023){Bailey}, {Bendek}, {Monacelli}, {Baker}, {Bedrosian}, {Cady}, {Douglas}, {Groff}, {Hildebrandt}, {Kasdin}, {Krist}, {Macintosh}, {Mennesson}, {Morrissey}, {Poberezhskiy}, {Subedi}, {Rhodes}, {Roberge}, {Ygouf}, {Zellem}, {Zhao}, \& {Zimmerman}}]{bailey2023}
{Bailey}, V.~P., {Bendek}, E., {Monacelli}, B., {et~al.} 2023, in Society of Photo-Optical Instrumentation Engineers (SPIE) Conference Series, Vol. 12680, Society of Photo-Optical Instrumentation Engineers (SPIE) Conference Series, 126800T, \dodoi{10.1117/12.2679036}

\bibitem[{{Baluev}(2008)}]{baluev2008}
{Baluev}, R.~V. 2008, \mnras, 389, 1375, \dodoi{10.1111/j.1365-2966.2008.13656.x}

\bibitem[{{Baluev}(2009)}]{baluev2009}
---. 2009, \mnras, 393, 969, \dodoi{10.1111/j.1365-2966.2008.14217.x}

\bibitem[{{Bardalez Gagliuffi} {et~al.}(2021){Bardalez Gagliuffi}, {Faherty}, {Li}, {Brandt}, {Williams}, {Brandt}, \& {Gelino}}]{bardalez2021}
{Bardalez Gagliuffi}, D.~C., {Faherty}, J.~K., {Li}, Y., {et~al.} 2021, \apjl, 922, L43, \dodoi{10.3847/2041-8213/ac382c}

\bibitem[{{Brown}(2015)}]{brown2015a}
{Brown}, R.~A. 2015, \apj, 799, 87, \dodoi{10.1088/0004-637X/799/1/87}

\bibitem[{{Burt} {et~al.}(2018){Burt}, {Holden}, {Wolfgang}, \& {Bouma}}]{burt2018}
{Burt}, J., {Holden}, B., {Wolfgang}, A., \& {Bouma}, L.~G. 2018, \aj, 156, 255, \dodoi{10.3847/1538-3881/aae697}

\bibitem[{{Butler} \& {Marcy}(1996)}]{butler1996a}
{Butler}, R.~P., \& {Marcy}, G.~W. 1996, \apjl, 464, L153, \dodoi{10.1086/310102}

\bibitem[{{Butler} {et~al.}(2003){Butler}, {Marcy}, {Vogt}, {Fischer}, {Henry}, {Laughlin}, \& {Wright}}]{butler2003}
{Butler}, R.~P., {Marcy}, G.~W., {Vogt}, S.~S., {et~al.} 2003, \apj, 582, 455, \dodoi{10.1086/344570}

\bibitem[{{Chaplin} {et~al.}(2019){Chaplin}, {Cegla}, {Watson}, {Davies}, \& {Ball}}]{chaplin2019}
{Chaplin}, W.~J., {Cegla}, H.~M., {Watson}, C.~A., {Davies}, G.~R., \& {Ball}, W.~H. 2019, \aj, 157, 163, \dodoi{10.3847/1538-3881/ab0c01}

\bibitem[{{Cloutier} {et~al.}(2018){Cloutier}, {Doyon}, {Bouchy}, \& {H{\'e}brard}}]{cloutier2018}
{Cloutier}, R., {Doyon}, R., {Bouchy}, F., \& {H{\'e}brard}, G. 2018, \aj, 156, 82, \dodoi{10.3847/1538-3881/aacea9}

\bibitem[{{Crass} {et~al.}(2021){Crass}, {Gaudi}, {Leifer}, {Beichman}, {Bender}, {Blackwood}, {Burt}, {Callas}, {Cegla}, {Diddams}, {Dumusque}, {Eastman}, {Ford}, {Fulton}, {Gibson}, {Halverson}, {Haywood}, {Hearty}, {Howard}, {Latham}, {L{\"o}hner-B{\"o}ttcher}, {Mamajek}, {Mortier}, {Newman}, {Plavchan}, {Quirrenbach}, {Reiners}, {Robertson}, {Roy}, {Schwab}, {Seifahrt}, {Szentgyorgyi}, {Terrien}, {Teske}, {Thompson}, \& {Vasisht}}]{crass2021}
{Crass}, J., {Gaudi}, B.~S., {Leifer}, S., {et~al.} 2021, arXiv e-prints, arXiv:2107.14291, \dodoi{10.48550/arXiv.2107.14291}

\bibitem[{{Dalba} {et~al.}(2021){Dalba}, {Kane}, {Howell}, {Horch}, {Li}, {Hirsch}, {Burt}, {Brandt}, {Mo{\v{c}}nik}, {Henry}, {Everett}, {Rosenthal}, \& {Howard}}]{dalba2021}
{Dalba}, P.~A., {Kane}, S.~R., {Howell}, S.~B., {et~al.} 2021, \aj, 161, 123, \dodoi{10.3847/1538-3881/abd6ed}

\bibitem[{{Dragomir} {et~al.}(2020){Dragomir}, {Harris}, {Pepper}, {Barclay}, {Villanueva}, {Ricker}, {Vanderspek}, {Latham}, {Seager}, {Winn}, {Jenkins}, {Ciardi}, {Furesz}, {Henze}, {Mireles}, {Morgan}, {Quintana}, {Ting}, \& {Yahalomi}}]{dragomir2020a}
{Dragomir}, D., {Harris}, M., {Pepper}, J., {et~al.} 2020, \aj, 159, 219, \dodoi{10.3847/1538-3881/ab845d}

\bibitem[{{Dumusque} {et~al.}(2011){Dumusque}, {Udry}, {Lovis}, {Santos}, \& {Monteiro}}]{dumusque2011}
{Dumusque}, X., {Udry}, S., {Lovis}, C., {Santos}, N.~C., \& {Monteiro}, M.~J.~P.~F.~G. 2011, \aap, 525, A140, \dodoi{10.1051/0004-6361/201014097}

\bibitem[{{Feng} {et~al.}(2022){Feng}, {Butler}, {Vogt}, {Clement}, {Tinney}, {Cui}, {Aizawa}, {Jones}, {Bailey}, {Burt}, {Carter}, {Crane}, {Flammini Dotti}, {Holden}, {Ma}, {Ogihara}, {Oppenheimer}, {O'Toole}, {Shectman}, {Wittenmyer}, {Wang}, {Wright}, \& {Xuan}}]{feng2022}
{Feng}, F., {Butler}, R.~P., {Vogt}, S.~S., {et~al.} 2022, \apjs, 262, 21, \dodoi{10.3847/1538-4365/ac7e57}

\bibitem[{{Fischer} {et~al.}(2002){Fischer}, {Marcy}, {Butler}, {Laughlin}, \& {Vogt}}]{fischer2002}
{Fischer}, D.~A., {Marcy}, G.~W., {Butler}, R.~P., {Laughlin}, G., \& {Vogt}, S.~S. 2002, \apj, 564, 1028, \dodoi{10.1086/324336}

\bibitem[{{Fischer} {et~al.}(2016){Fischer}, {Anglada-Escude}, {Arriagada}, {Baluev}, {Bean}, {Bouchy}, {Buchhave}, {Carroll}, {Chakraborty}, {Crepp}, {Dawson}, {Diddams}, {Dumusque}, {Eastman}, {Endl}, {Figueira}, {Ford}, {Foreman-Mackey}, {Fournier}, {F{\H{u}}r{\'e}sz}, {Gaudi}, {Gregory}, {Grundahl}, {Hatzes}, {H{\'e}brard}, {Herrero}, {Hogg}, {Howard}, {Johnson}, {Jorden}, {Jurgenson}, {Latham}, {Laughlin}, {Loredo}, {Lovis}, {Mahadevan}, {McCracken}, {Pepe}, {Perez}, {Phillips}, {Plavchan}, {Prato}, {Quirrenbach}, {Reiners}, {Robertson}, {Santos}, {Sawyer}, {Segransan}, {Sozzetti}, {Steinmetz}, {Szentgyorgyi}, {Udry}, {Valenti}, {Wang}, {Wittenmyer}, \& {Wright}}]{fischer2016}
{Fischer}, D.~A., {Anglada-Escude}, G., {Arriagada}, P., {et~al.} 2016, \pasp, 128, 066001, \dodoi{10.1088/1538-3873/128/964/066001}

\bibitem[{{Fulton} {et~al.}(2018){Fulton}, {Petigura}, {Blunt}, \& {Sinukoff}}]{fulton2018a}
{Fulton}, B.~J., {Petigura}, E.~A., {Blunt}, S., \& {Sinukoff}, E. 2018, \pasp, 130, 044504, \dodoi{10.1088/1538-3873/aaaaa8}

\bibitem[{{Gaia Collaboration} {et~al.}(2018){Gaia Collaboration}, {Brown}, {Vallenari}, {Prusti}, {de Bruijne}, {Babusiaux}, {Bailer-Jones}, {Biermann}, {Evans}, {Eyer}, {Jansen}, {Jordi}, {Klioner}, {Lammers}, {Lindegren}, {Luri}, {Mignard}, {Panem}, {Pourbaix}, {Randich}, {Sartoretti}, {Siddiqui}, {Soubiran}, {van Leeuwen}, {Walton}, {Arenou}, {Bastian}, {Cropper}, {Drimmel}, {Katz}, {Lattanzi}, {Bakker}, {Cacciari}, {Casta{\~n}eda}, {Chaoul}, {Cheek}, {De Angeli}, {Fabricius}, {Guerra}, {Holl}, {Masana}, {Messineo}, {Mowlavi}, {Nienartowicz}, {Panuzzo}, {Portell}, {Riello}, {Seabroke}, {Tanga}, {Th{\'e}venin}, {Gracia-Abril}, {Comoretto}, {Garcia-Reinaldos}, {Teyssier}, {Altmann}, {Andrae}, {Audard}, {Bellas-Velidis}, {Benson}, {Berthier}, {Blomme}, {Burgess}, {Busso}, {Carry}, {Cellino}, {Clementini}, {Clotet}, {Creevey}, {Davidson}, {De Ridder}, {Delchambre}, {Dell'Oro}, {Ducourant}, {Fern{\'a}ndez-Hern{\'a}ndez}, {Fouesneau}, {Fr{\'e}mat}, {Galluccio}, {Garc{\'\i}a-Torres},
  {Gonz{\'a}lez-N{\'u}{\~n}ez}, {Gonz{\'a}lez-Vidal}, {Gosset}, {Guy}, {Halbwachs}, {Hambly}, {Harrison}, {Hern{\'a}ndez}, {Hestroffer}, {Hodgkin}, {Hutton}, {Jasniewicz}, {Jean-Antoine-Piccolo}, {Jordan}, {Korn}, {Krone-Martins}, {Lanzafame}, {Lebzelter}, {L{\"o}ffler}, {Manteiga}, {Marrese}, {Mart{\'\i}n-Fleitas}, {Moitinho}, {Mora}, {Muinonen}, {Osinde}, {Pancino}, {Pauwels}, {Petit}, {Recio-Blanco}, {Richards}, {Rimoldini}, {Robin}, {Sarro}, {Siopis}, {Smith}, {Sozzetti}, {S{\"u}veges}, {Torra}, {van Reeven}, {Abbas}, {Abreu Aramburu}, {Accart}, {Aerts}, {Altavilla}, {{\'A}lvarez}, {Alvarez}, {Alves}, {Anderson}, {Andrei}, {Anglada Varela}, {Antiche}, {Antoja}, {Arcay}, {Astraatmadja}, {Bach}, {Baker}, {Balaguer-N{\'u}{\~n}ez}, {Balm}, {Barache}, {Barata}, {Barbato}, {Barblan}, {Barklem}, {Barrado}, {Barros}, {Barstow}, {Bartholom{\'e} Mu{\~n}oz}, {Bassilana}, {Becciani}, {Bellazzini}, {Berihuete}, {Bertone}, {Bianchi}, {Bienaym{\'e}}, {Blanco-Cuaresma}, {Boch}, {Boeche}, {Bombrun}, {Borrachero},
  {Bossini}, {Bouquillon}, {Bourda}, {Bragaglia}, {Bramante}, {Breddels}, {Bressan}, {Brouillet}, {Br{\"u}semeister}, {Brugaletta}, {Bucciarelli}, {Burlacu}, {Busonero}, {Butkevich}, {Buzzi}, {Caffau}, {Cancelliere}, {Cannizzaro}, {Cantat-Gaudin}, {Carballo}, {Carlucci}, {Carrasco}, {Casamiquela}, {Castellani}, {Castro-Ginard}, {Charlot}, {Chemin}, {Chiavassa}, {Cocozza}, {Costigan}, {Cowell}, {Crifo}, {Crosta}, {Crowley}, {Cuypers}, {Dafonte}, {Damerdji}, {Dapergolas}, {David}, {David}, {de Laverny}, {De Luise}, {De March}, {de Martino}, {de Souza}, {de Torres}, {Debosscher}, {del Pozo}, {Delbo}, {Delgado}, {Delgado}, {Di Matteo}, {Diakite}, {Diener}, {Distefano}, {Dolding}, {Drazinos}, {Dur{\'a}n}, {Edvardsson}, {Enke}, {Eriksson}, {Esquej}, {Eynard Bontemps}, {Fabre}, {Fabrizio}, {Faigler}, {Falc{\~a}o}, {Farr{\`a}s Casas}, {Federici}, {Fedorets}, {Fernique}, {Figueras}, {Filippi}, {Findeisen}, {Fonti}, {Fraile}, {Fraser}, {Fr{\'e}zouls}, {Gai}, {Galleti}, {Garabato}, {Garc{\'\i}a-Sedano}, {Garofalo},
  {Garralda}, {Gavel}, {Gavras}, {Gerssen}, {Geyer}, {Giacobbe}, {Gilmore}, {Girona}, {Giuffrida}, {Glass}, {Gomes}, {Granvik}, {Gueguen}, {Guerrier}, {Guiraud}, {Guti{\'e}rrez-S{\'a}nchez}, {Haigron}, {Hatzidimitriou}, {Hauser}, {Haywood}, {Heiter}, {Helmi}, {Heu}, {Hilger}, {Hobbs}, {Hofmann}, {Holland}, {Huckle}, {Hypki}, {Icardi}, {Jan{\ss}en}, {Jevardat de Fombelle}, {Jonker}, {Juh{\'a}sz}, {Julbe}, {Karampelas}, {Kewley}, {Klar}, {Kochoska}, {Kohley}, {Kolenberg}, {Kontizas}, {Kontizas}, {Koposov}, {Kordopatis}, {Kostrzewa-Rutkowska}, {Koubsky}, {Lambert}, {Lanza}, {Lasne}, {Lavigne}, {Le Fustec}, {Le Poncin-Lafitte}, {Lebreton}, {Leccia}, {Leclerc}, {Lecoeur-Taibi}, {Lenhardt}, {Leroux}, {Liao}, {Licata}, {Lindstr{\o}m}, {Lister}, {Livanou}, {Lobel}, {L{\'o}pez}, {Managau}, {Mann}, {Mantelet}, {Marchal}, {Marchant}, {Marconi}, {Marinoni}, {Marschalk{\'o}}, {Marshall}, {Martino}, {Marton}, {Mary}, {Massari}, {Matijevi{\v{c}}}, {Mazeh}, {McMillan}, {Messina}, {Michalik}, {Millar}, {Molina}, {Molinaro},
  {Moln{\'a}r}, {Montegriffo}, {Mor}, {Morbidelli}, {Morel}, {Morris}, {Mulone}, {Muraveva}, {Musella}, {Nelemans}, {Nicastro}, {Noval}, {O'Mullane}, {Ord{\'e}novic}, {Ord{\'o}{\~n}ez-Blanco}, {Osborne}, {Pagani}, {Pagano}, {Pailler}, {Palacin}, {Palaversa}, {Panahi}, {Pawlak}, {Piersimoni}, {Pineau}, {Plachy}, {Plum}, {Poggio}, {Poujoulet}, {Pr{\v{s}}a}, {Pulone}, {Racero}, {Ragaini}, {Rambaux}, {Ramos-Lerate}, {Regibo}, {Reyl{\'e}}, {Riclet}, {Ripepi}, {Riva}, {Rivard}, {Rixon}, {Roegiers}, {Roelens}, {Romero-G{\'o}mez}, {Rowell}, {Royer}, {Ruiz-Dern}, {Sadowski}, {Sagrist{\`a} Sell{\'e}s}, {Sahlmann}, {Salgado}, {Salguero}, {Sanna}, {Santana-Ros}, {Sarasso}, {Savietto}, {Schultheis}, {Sciacca}, {Segol}, {Segovia}, {S{\'e}gransan}, {Shih}, {Siltala}, {Silva}, {Smart}, {Smith}, {Solano}, {Solitro}, {Sordo}, {Soria Nieto}, {Souchay}, {Spagna}, {Spoto}, {Stampa}, {Steele}, {Steidelm{\"u}ller}, {Stephenson}, {Stoev}, {Suess}, {Surdej}, {Szabados}, {Szegedi-Elek}, {Tapiador}, {Taris}, {Tauran}, {Taylor},
  {Teixeira}, {Terrett}, {Teyssand ier}, {Thuillot}, {Titarenko}, {Torra Clotet}, {Turon}, {Ulla}, {Utrilla}, {Uzzi}, {Vaillant}, {Valentini}, {Valette}, {van Elteren}, {Van Hemelryck}, {van Leeuwen}, {Vaschetto}, {Vecchiato}, {Veljanoski}, {Viala}, {Vicente}, {Vogt}, {von Essen}, {Voss}, {Votruba}, {Voutsinas}, {Walmsley}, {Weiler}, {Wertz}, {Wevers}, {Wyrzykowski}, {Yoldas}, {{\v{Z}}erjal}, {Ziaeepour}, {Zorec}, {Zschocke}, {Zucker}, {Zurbach}, \& {Zwitter}}]{gaia2018}
{Gaia Collaboration}, {Brown}, A.~G.~A., {Vallenari}, A., {et~al.} 2018, \aap, 616, A1, \dodoi{10.1051/0004-6361/201833051}

\bibitem[{{Gaudi} {et~al.}(2020){Gaudi}, {Seager}, {Mennesson}, {Kiessling}, {Warfield}, {Cahoy}, {Clarke}, {Domagal-Goldman}, {Feinberg}, {Guyon}, {Kasdin}, {Mawet}, {Plavchan}, {Robinson}, {Rogers}, {Scowen}, {Somerville}, {Stapelfeldt}, {Stark}, {Stern}, {Turnbull}, {Amini}, {Kuan}, {Martin}, {Morgan}, {Redding}, {Stahl}, {Webb}, {Alvarez-Salazar}, {Arnold}, {Arya}, {Balasubramanian}, {Baysinger}, {Bell}, {Below}, {Benson}, {Blais}, {Booth}, {Bourgeois}, {Bradford}, {Brewer}, {Brooks}, {Cady}, {Caldwell}, {Calvet}, {Carr}, {Chan}, {Cormarkovic}, {Coste}, {Cox}, {Danner}, {Davis}, {Dewell}, {Dorsett}, {Dunn}, {East}, {Effinger}, {Eng}, {Freebury}, {Garcia}, {Gaskin}, {Greene}, {Hennessy}, {Hilgemann}, {Hood}, {Holota}, {Howe}, {Huang}, {Hull}, {Hunt}, {Hurd}, {Johnson}, {Kissil}, {Knight}, {Kolenz}, {Kraus}, {Krist}, {Li}, {Lisman}, {Mandic}, {Mann}, {Marchen}, {Marrese-Reading}, {McCready}, {McGown}, {Missun}, {Miyaguchi}, {Moore}, {Nemati}, {Nikzad}, {Nissen}, {Novicki}, {Perrine}, {Pineda}, {Polanco},
  {Putnam}, {Qureshi}, {Richards}, {Eldorado Riggs}, {Rodgers}, {Rud}, {Saini}, {Scalisi}, {Scharf}, {Schulz}, {Serabyn}, {Sigrist}, {Sikkia}, {Singleton}, {Shaklan}, {Smith}, {Southerd}, {Stahl}, {Steeves}, {Sturges}, {Sullivan}, {Tang}, {Taras}, {Tesch}, {Therrell}, {Tseng}, {Valente}, {Van Buren}, {Villalvazo}, {Warwick}, {Webb}, {Westerhoff}, {Wofford}, {Wu}, {Woo}, {Wood}, {Ziemer}, {Arney}, {Anderson}, {Ma{\'\i}z-Apell{\'a}niz}, {Bartlett}, {Belikov}, {Bendek}, {Cenko}, {Douglas}, {Dulz}, {Evans}, {Faramaz}, {Feng}, {Ferguson}, {Follette}, {Ford}, {Garc{\'\i}a}, {Geha}, {Gelino}, {G{\"o}tberg}, {Hildebrandt}, {Hu}, {Jahnke}, {Kennedy}, {Kreidberg}, {Isella}, {Lopez}, {Marchis}, {Macri}, {Marley}, {Matzko}, {Mazoyer}, {McCandliss}, {Meshkat}, {Mordasini}, {Morris}, {Nielsen}, {Newman}, {Petigura}, {Postman}, {Reines}, {Roberge}, {Roederer}, {Ruane}, {Schwieterman}, {Sirbu}, {Spalding}, {Teplitz}, {Tumlinson}, {Turner}, {Werk}, {Wofford}, {Wyatt}, {Young}, \& {Zellem}}]{gaudi2020}
{Gaudi}, B.~S., {Seager}, S., {Mennesson}, B., {et~al.} 2020, arXiv e-prints, arXiv:2001.06683, \dodoi{10.48550/arXiv.2001.06683}

\bibitem[{{Gibson} {et~al.}(2024){Gibson}, {Howard}, {Rider}, {Halverson}, {Roy}, {Baker}, {Edelstein}, {Smith}, {Fulton}, {Walawender}, {Brodheim}, {Brown}, {Chan}, {Dai}, {Deich}, {Gottschalk}, {Grillo}, {Hale}, {Hill}, {Holden}, {Householder}, {Isaacson}, {Ishikawa}, {Jelinsky}, {Kassis}, {Kaye}, {Laher}, {Lanclos}, {Lee}, {Lilley}, {McCarney}, {Miller}, {Payne}, {Petigura}, {Poppett}, {Raffanti}, {Rubenzahl}, {Sandford}, {Schwab}, {Shaum}, {Sirk}, {Smith}, {Thorne}, {Valliant}, {Vandenberg}, {Wang}, {Wishnow}, {Wold}, {Yeh}, {Baca}, {Beichman}, {Berriman}, {Brown}, {Casey}, {Chin}, {Chong}, {Cowley}, {Devenot}, {Elwir}, {Finstad}, {Fraysse}, {James}, {Jhoti}, {Killian}, {Levine}, {Li}, {Marin}, {Milner}, {Nance}, {O'Hanlon}, {Orr}, {Ortiz-Soto}, {Payne}, {Pember}, {Raskin}, {Savage}, {Seifahrt}, {Smith}, {Storesund}, {St{\"u}rmer}, {Suominen}, {Tehero}, {Von Boeckmann}, {Wages}, {Weisfeiler}, {Wilcox}, {Wizinowich}, \& {Wolfenberger}}]{gibson2024}
{Gibson}, S.~R., {Howard}, A.~W., {Rider}, K., {et~al.} 2024, in Society of Photo-Optical Instrumentation Engineers (SPIE) Conference Series, Vol. 13096, Ground-based and Airborne Instrumentation for Astronomy X, ed. J.~J. {Bryant}, K.~{Motohara}, \& J.~R.~D. {Vernet}, 1309609, \dodoi{10.1117/12.3017841}

\bibitem[{{Gregory} \& {Fischer}(2010)}]{gregory2010}
{Gregory}, P.~C., \& {Fischer}, D.~A. 2010, \mnras, 403, 731, \dodoi{10.1111/j.1365-2966.2009.16233.x}

\bibitem[{{Gupta} \& {Bedell}(2024)}]{gupta2024}
{Gupta}, A.~F., \& {Bedell}, M. 2024, \aj, 168, 29, \dodoi{10.3847/1538-3881/ad4ce6}

\bibitem[{{Hall} {et~al.}(2018){Hall}, {Thompson}, {Handley}, \& {Queloz}}]{hall2018}
{Hall}, R.~D., {Thompson}, S.~J., {Handley}, W., \& {Queloz}, D. 2018, \mnras, 479, 2968, \dodoi{10.1093/mnras/sty1464}

\bibitem[{{Harada} {et~al.}(2024{\natexlab{a}}){Harada}, {Dressing}, {Kane}, \& {Ardestani}}]{harada2024b}
{Harada}, C.~K., {Dressing}, C.~D., {Kane}, S.~R., \& {Ardestani}, B.~A. 2024{\natexlab{a}}, \apjs, 272, 30, \dodoi{10.3847/1538-4365/ad3e81}

\bibitem[{{Harada} {et~al.}(2024{\natexlab{b}}){Harada}, {Dressing}, {Kane}, {Blunt}, {Dietrich}, {Hinkel}, {Li}, {Mamajek}, {Rice}, {Tuchow}, {Turtelboom}, \& {Wittenmyer}}]{harada2024c}
{Harada}, C.~K., {Dressing}, C.~D., {Kane}, S.~R., {et~al.} 2024{\natexlab{b}}, arXiv e-prints, arXiv:2409.10679, \dodoi{10.48550/arXiv.2409.10679}

\bibitem[{Harris {et~al.}(2020)Harris, Millman, van~der Walt, Gommers, Virtanen, Cournapeau, Wieser, Taylor, Berg, Smith, Kern, Picus, Hoyer, van Kerkwijk, Brett, Haldane, del R{\'{i}}o, Wiebe, Peterson, G{\'{e}}rard-Marchant, Sheppard, Reddy, Weckesser, Abbasi, Gohlke, \& Oliphant}]{harris2020}
Harris, C.~R., Millman, K.~J., van~der Walt, S.~J., {et~al.} 2020, Nature, 585, 357, \dodoi{10.1038/s41586-020-2649-2}

\bibitem[{{Hill} {et~al.}(2023){Hill}, {Bott}, {Dalba}, {Fetherolf}, {Kane}, {Kopparapu}, {Li}, \& {Ostberg}}]{hill2023}
{Hill}, M.~L., {Bott}, K., {Dalba}, P.~A., {et~al.} 2023, \aj, 165, 34, \dodoi{10.3847/1538-3881/aca1c0}

\bibitem[{{Hill} {et~al.}(2018){Hill}, {Kane}, {Seperuelo Duarte}, {Kopparapu}, {Gelino}, \& {Wittenmyer}}]{hill2018}
{Hill}, M.~L., {Kane}, S.~R., {Seperuelo Duarte}, E., {et~al.} 2018, \apj, 860, 67, \dodoi{10.3847/1538-4357/aac384}

\bibitem[{{Hunter}(2007)}]{hunter2007}
{Hunter}, J.~D. 2007, Computing in Science and Engineering, 9, 90, \dodoi{10.1109/MCSE.2007.55}

\bibitem[{{Jones} {et~al.}(2010){Jones}, {Butler}, {Tinney}, {O'Toole}, {Wittenmyer}, {Henry}, {Meschiari}, {Vogt}, {Rivera}, {Laughlin}, {Carter}, {Bailey}, \& {Jenkins}}]{jones2010}
{Jones}, H.~R.~A., {Butler}, R.~P., {Tinney}, C.~G., {et~al.} 2010, \mnras, 403, 1703, \dodoi{10.1111/j.1365-2966.2009.16232.x}

\bibitem[{{Jurgenson} {et~al.}(2016){Jurgenson}, {Fischer}, {McCracken}, {Sawyer}, {Szymkowiak}, {Davis}, {Muller}, \& {Santoro}}]{jurgenson2016}
{Jurgenson}, C., {Fischer}, D., {McCracken}, T., {et~al.} 2016, in Society of Photo-Optical Instrumentation Engineers (SPIE) Conference Series, Vol. 9908, Ground-based and Airborne Instrumentation for Astronomy VI, ed. C.~J. {Evans}, L.~{Simard}, \& H.~{Takami}, 99086T, \dodoi{10.1117/12.2233002}

\bibitem[{{Kane}(2007)}]{kane2007b}
{Kane}, S.~R. 2007, \mnras, 380, 1488, \dodoi{10.1111/j.1365-2966.2007.12144.x}

\bibitem[{{Kane}(2013)}]{kane2013c}
---. 2013, \apj, 766, 10, \dodoi{10.1088/0004-637X/766/1/10}

\bibitem[{{Kane} \& {Burt}(2024)}]{kane2024e}
{Kane}, S.~R., \& {Burt}, J.~A. 2024, \aj, 168, 279, \dodoi{10.3847/1538-3881/ad8a68}

\bibitem[{{Kane} \& {Gelino}(2012)}]{kane2012a}
{Kane}, S.~R., \& {Gelino}, D.~M. 2012, \pasp, 124, 323, \dodoi{10.1086/665271}

\bibitem[{{Kane} {et~al.}(2024){Kane}, {Li}, {Turnbull}, {Dressing}, \& {Harada}}]{kane2024d}
{Kane}, S.~R., {Li}, Z., {Turnbull}, M.~C., {Dressing}, C.~D., \& {Harada}, C.~K. 2024, arXiv e-prints, arXiv:2408.00263, \dodoi{10.48550/arXiv.2408.00263}

\bibitem[{{Kane} {et~al.}(2009){Kane}, {Mahadevan}, {von Braun}, {Laughlin}, \& {Ciardi}}]{kane2009c}
{Kane}, S.~R., {Mahadevan}, S., {von Braun}, K., {Laughlin}, G., \& {Ciardi}, D.~R. 2009, \pasp, 121, 1386, \dodoi{10.1086/648564}

\bibitem[{{Kane} {et~al.}(2018){Kane}, {Meshkat}, \& {Turnbull}}]{kane2018c}
{Kane}, S.~R., {Meshkat}, T., \& {Turnbull}, M.~C. 2018, \aj, 156, 267, \dodoi{10.3847/1538-3881/aae981}

\bibitem[{{Kane} {et~al.}(2016){Kane}, {Hill}, {Kasting}, {Kopparapu}, {Quintana}, {Barclay}, {Batalha}, {Borucki}, {Ciardi}, {Haghighipour}, {Hinkel}, {Kaltenegger}, {Selsis}, \& {Torres}}]{kane2016c}
{Kane}, S.~R., {Hill}, M.~L., {Kasting}, J.~F., {et~al.} 2016, \apj, 830, 1, \dodoi{10.3847/0004-637X/830/1/1}

\bibitem[{{Kane} {et~al.}(2019){Kane}, {Dalba}, {Li}, {Horch}, {Hirsch}, {Horner}, {Wittenmyer}, {Howell}, {Everett}, {Butler}, {Tinney}, {Carter}, {Wright}, {Jones}, {Bailey}, \& {O'Toole}}]{kane2019b}
{Kane}, S.~R., {Dalba}, P.~A., {Li}, Z., {et~al.} 2019, \aj, 157, 252, \dodoi{10.3847/1538-3881/ab1ddf}

\bibitem[{{Kane} {et~al.}(2021){Kane}, {Bean}, {Campante}, {Dalba}, {Fetherolf}, {Mocnik}, {Ostberg}, {Pepper}, {Simpson}, {Turnbull}, {Ricker}, {Vanderspek}, {Latham}, {Seager}, {Winn}, {Jenkins}, {Huber}, \& {Chaplin}}]{kane2021b}
{Kane}, S.~R., {Bean}, J.~L., {Campante}, T.~L., {et~al.} 2021, \pasp, 133, 014402, \dodoi{10.1088/1538-3873/abc610}

\bibitem[{{Kasting} {et~al.}(1993){Kasting}, {Whitmire}, \& {Reynolds}}]{kasting1993a}
{Kasting}, J.~F., {Whitmire}, D.~P., \& {Reynolds}, R.~T. 1993, \icarus, 101, 108, \dodoi{10.1006/icar.1993.1010}

\bibitem[{{Kopparapu} {et~al.}(2014){Kopparapu}, {Ramirez}, {SchottelKotte}, {Kasting}, {Domagal-Goldman}, \& {Eymet}}]{kopparapu2014}
{Kopparapu}, R.~K., {Ramirez}, R.~M., {SchottelKotte}, J., {et~al.} 2014, \apjl, 787, L29, \dodoi{10.1088/2041-8205/787/2/L29}

\bibitem[{{Kopparapu} {et~al.}(2013){Kopparapu}, {Ramirez}, {Kasting}, {Eymet}, {Robinson}, {Mahadevan}, {Terrien}, {Domagal-Goldman}, {Meadows}, \& {Deshpande}}]{kopparapu2013a}
{Kopparapu}, R.~K., {Ramirez}, R., {Kasting}, J.~F., {et~al.} 2013, \apj, 765, 131, \dodoi{10.1088/0004-637X/765/2/131}

\bibitem[{{Laliotis} {et~al.}(2023){Laliotis}, {Burt}, {Mamajek}, {Li}, {Perdelwitz}, {Zhao}, {Butler}, {Holden}, {Rosenthal}, {Fulton}, {Feng}, {Kane}, {Bailey}, {Carter}, {Crane}, {Furlan}, {Gnilka}, {Howell}, {Laughlin}, {Shectman}, {Teske}, {Tinney}, {Vogt}, {Wang}, \& {Wittenmyer}}]{laliotis2023}
{Laliotis}, K., {Burt}, J.~A., {Mamajek}, E.~E., {et~al.} 2023, \aj, 165, 176, \dodoi{10.3847/1538-3881/acc067}

\bibitem[{{Lam} {et~al.}(2024){Lam}, {Bedell}, {Zhao}, {Gupta}, \& {Ballard}}]{lam2024}
{Lam}, C., {Bedell}, M., {Zhao}, L.~L., {Gupta}, A.~F., \& {Ballard}, S.~A. 2024, \aj, 168, 200, \dodoi{10.3847/1538-3881/ad739b}

\bibitem[{{Li} {et~al.}(2021){Li}, {Hildebrandt}, {Kane}, {Zimmerman}, {Girard}, {Gonzalez-Quiles}, \& {Turnbull}}]{li2021}
{Li}, Z., {Hildebrandt}, S.~R., {Kane}, S.~R., {et~al.} 2021, \aj, 162, 9, \dodoi{10.3847/1538-3881/abf831}

\bibitem[{{Lucy} \& {Sweeney}(1971)}]{lucy1971}
{Lucy}, L.~B., \& {Sweeney}, M.~A. 1971, \aj, 76, 544, \dodoi{10.1086/111159}

\bibitem[{{Luhn} {et~al.}(2023){Luhn}, {Ford}, {Guo}, {Gilbertson}, {Newman}, {Plavchan}, {Burt}, {Teske}, \& {Gupta}}]{luhn2023}
{Luhn}, J.~K., {Ford}, E.~B., {Guo}, Z., {et~al.} 2023, \aj, 165, 98, \dodoi{10.3847/1538-3881/acad08}

\bibitem[{{Marmier} {et~al.}(2013){Marmier}, {S{\'e}gransan}, {Udry}, {Mayor}, {Pepe}, {Queloz}, {Lovis}, {Naef}, {Santos}, {Alonso}, {Alves}, {Berthet}, {Chazelas}, {Demory}, {Dumusque}, {Eggenberger}, {Figueira}, {Gillon}, {Hagelberg}, {Lendl}, {Mardling}, {M{\'e}gevand}, {Neveu}, {Sahlmann}, {Sosnowska}, {Tewes}, \& {Triaud}}]{marmier2013}
{Marmier}, M., {S{\'e}gransan}, D., {Udry}, S., {et~al.} 2013, \aap, 551, A90, \dodoi{10.1051/0004-6361/201219639}

\bibitem[{{Meunier} {et~al.}(2023){Meunier}, {Pous}, {Sulis}, {Mary}, \& {Lagrange}}]{meunier2023}
{Meunier}, N., {Pous}, R., {Sulis}, S., {Mary}, D., \& {Lagrange}, A.~M. 2023, \aap, 676, A82, \dodoi{10.1051/0004-6361/202346218}

\bibitem[{{Naef} {et~al.}(2004){Naef}, {Mayor}, {Beuzit}, {Perrier}, {Queloz}, {Sivan}, \& {Udry}}]{naef2004}
{Naef}, D., {Mayor}, M., {Beuzit}, J.~L., {et~al.} 2004, \aap, 414, 351, \dodoi{10.1051/0004-6361:20034091}

\bibitem[{{National Academies of Sciences, Engineering, and Medicine}(2021)}]{nas2021}
{National Academies of Sciences, Engineering, and Medicine}. 2021, {Pathways to Discovery in Astronomy and Astrophysics for the 2020s}, \dodoi{10.17226/26141}

\bibitem[{{Newman} {et~al.}(2023){Newman}, {Plavchan}, {Burt}, {Teske}, {Mamajek}, {Leifer}, {Gaudi}, {Blackwood}, \& {Morgan}}]{newman2023}
{Newman}, P.~D., {Plavchan}, P., {Burt}, J.~A., {et~al.} 2023, \aj, 165, 151, \dodoi{10.3847/1538-3881/acad07}

\bibitem[{Pedregosa {et~al.}(2011)Pedregosa, Varoquaux, Gramfort, Michel, Thirion, Grisel, Blondel, Prettenhofer, Weiss, Dubourg, Vanderplas, Passos, Cournapeau, Brucher, Perrot, \& Duchesnay}]{pedregosa2011}
Pedregosa, F., Varoquaux, G., Gramfort, A., {et~al.} 2011, Journal of Machine Learning Research, 12, 2825

\bibitem[{{Pepe} {et~al.}(2021){Pepe}, {Cristiani}, {Rebolo}, {Santos}, {Dekker}, {Cabral}, {Di Marcantonio}, {Figueira}, {Lo Curto}, {Lovis}, {Mayor}, {M{\'e}gevand}, {Molaro}, {Riva}, {Zapatero Osorio}, {Amate}, {Manescau}, {Pasquini}, {Zerbi}, {Adibekyan}, {Abreu}, {Affolter}, {Alibert}, {Aliverti}, {Allart}, {Allende Prieto}, {{\'A}lvarez}, {Alves}, {Avila}, {Baldini}, {Bandy}, {Barros}, {Benz}, {Bianco}, {Borsa}, {Bourrier}, {Bouchy}, {Broeg}, {Calderone}, {Cirami}, {Coelho}, {Conconi}, {Coretti}, {Cumani}, {Cupani}, {D'Odorico}, {Damasso}, {Deiries}, {Delabre}, {Demangeon}, {Dumusque}, {Ehrenreich}, {Faria}, {Fragoso}, {Genolet}, {Genoni}, {G{\'e}nova Santos}, {Gonz{\'a}lez Hern{\'a}ndez}, {Hughes}, {Iwert}, {Kerber}, {Knudstrup}, {Landoni}, {Lavie}, {Lillo-Box}, {Lizon}, {Maire}, {Martins}, {Mehner}, {Micela}, {Modigliani}, {Monteiro}, {Monteiro}, {Moschetti}, {Murphy}, {Nunes}, {Oggioni}, {Oliveira}, {Oshagh}, {Pall{\'e}}, {Pariani}, {Poretti}, {Rasilla}, {Rebord{\~a}o}, {Redaelli}, {Santana Tschudi},
  {Santin}, {Santos}, {S{\'e}gransan}, {Schmidt}, {Segovia}, {Sosnowska}, {Sozzetti}, {Sousa}, {Span{\`o}}, {Su{\'a}rez Mascare{\~n}o}, {Tabernero}, {Tenegi}, {Udry}, \& {Zanutta}}]{pepe2021}
{Pepe}, F., {Cristiani}, S., {Rebolo}, R., {et~al.} 2021, \aap, 645, A96, \dodoi{10.1051/0004-6361/202038306}

\bibitem[{{Riggs} {et~al.}(2021){Riggs}, {Bailey}, {Moody}, {Sidick}, {Balasubramanian}, {Moore}, {Wilson}, {Ruane}, {Sirbu}, {Gersh-Range}, {Trauger}, {Mennesson}, {Siegler}, {Bendek}, {Groff}, {Zimmerman}, {Debes}, {Basinger}, \& {Kasdin}}]{riggs2021}
{Riggs}, A.~J.~E., {Bailey}, V., {Moody}, D.~C., {et~al.} 2021, in Society of Photo-Optical Instrumentation Engineers (SPIE) Conference Series, Vol. 11823, Techniques and Instrumentation for Detection of Exoplanets X, ed. S.~B. {Shaklan} \& G.~J. {Ruane}, 118231Y, \dodoi{10.1117/12.2598599}

\bibitem[{{Rosenthal} {et~al.}(2021){Rosenthal}, {Fulton}, {Hirsch}, {Isaacson}, {Howard}, {Dedrick}, {Sherstyuk}, {Blunt}, {Petigura}, {Knutson}, {Behmard}, {Chontos}, {Crepp}, {Crossfield}, {Dalba}, {Fischer}, {Henry}, {Kane}, {Kosiarek}, {Marcy}, {Rubenzahl}, {Weiss}, \& {Wright}}]{rosenthal2021}
{Rosenthal}, L.~J., {Fulton}, B.~J., {Hirsch}, L.~A., {et~al.} 2021, \apjs, 255, 8, \dodoi{10.3847/1538-4365/abe23c}

\bibitem[{{Savransky} {et~al.}(2024){Savransky}, {Bailey}, {Wolff}, {Millar-Blanchaer}, {Wang}, {Altinier}, {Anche}, {Baudoz}, {Biller}, {Blunt}, {Brandner}, {Brinjikji}, {Carri{\'o}n-Gonz{\'a}lez}, {Chavez}, {Choquet}, {Doelman}, {Girard}, {Greenbaum}, {Hasler}, {Hom}, {Ingalls}, {Kane}, {Kasdin}, {Krause}, {Kuzuhara}, {Lau}, {Li}, {Livingston}, {Lowrance}, {Ludwick}, {Macintosh}, {Mamajek}, {Marley}, {Mazoyer}, {Mennesson}, {Mizuki}, {Moran}, {Murakami}, {Nishikawa}, {Noel}, {Pueyo}, {Hildebrandt Rafels}, {Rhodes}, {Robinson}, {De Rosa}, {Samland}, {Schragal}, {Schreiber}, {Sobeck}, {Stapelfeldt}, {Tamura}, {Uyajma}, {Vigan}, {Woodland}, {Ygouf}, {Yoneta}, {Zellem}, \& {Zimmerman}}]{savransky2024}
{Savransky}, D., {Bailey}, V.~P., {Wolff}, S.~G., {et~al.} 2024, in Society of Photo-Optical Instrumentation Engineers (SPIE) Conference Series, Vol. 13092, Space Telescopes and Instrumentation 2024: Optical, Infrared, and Millimeter Wave, ed. L.~E. {Coyle}, S.~{Matsuura}, \& M.~D. {Perrin}, 130921I, \dodoi{10.1117/12.3020514}

\bibitem[{{Schwab} {et~al.}(2016){Schwab}, {Rakich}, {Gong}, {Mahadevan}, {Halverson}, {Roy}, {Terrien}, {Robertson}, {Hearty}, {Levi}, {Monson}, {Wright}, {McElwain}, {Bender}, {Blake}, {St{\"u}rmer}, {Gurevich}, {Chakraborty}, \& {Ramsey}}]{schwab2016}
{Schwab}, C., {Rakich}, A., {Gong}, Q., {et~al.} 2016, in Society of Photo-Optical Instrumentation Engineers (SPIE) Conference Series, Vol. 9908, Ground-based and Airborne Instrumentation for Astronomy VI, ed. C.~J. {Evans}, L.~{Simard}, \& H.~{Takami}, 99087H, \dodoi{10.1117/12.2234411}

\bibitem[{Servén \& Brummitt(2018)}]{serven2018}
Servén, D., \& Brummitt, C. 2018, pyGAM: Generalized Additive Models in Python, \dodoi{10.5281/zenodo.1208723}

\bibitem[{{Spergel} {et~al.}(2015){Spergel}, {Gehrels}, {Baltay}, {Bennett}, {Breckinridge}, {Donahue}, {Dressler}, {Gaudi}, {Greene}, {Guyon}, {Hirata}, {Kalirai}, {Kasdin}, {Macintosh}, {Moos}, {Perlmutter}, {Postman}, {Rauscher}, {Rhodes}, {Wang}, {Weinberg}, {Benford}, {Hudson}, {Jeong}, {Mellier}, {Traub}, {Yamada}, {Capak}, {Colbert}, {Masters}, {Penny}, {Savransky}, {Stern}, {Zimmerman}, {Barry}, {Bartusek}, {Carpenter}, {Cheng}, {Content}, {Dekens}, {Demers}, {Grady}, {Jackson}, {Kuan}, {Kruk}, {Melton}, {Nemati}, {Parvin}, {Poberezhskiy}, {Peddie}, {Ruffa}, {Wallace}, {Whipple}, {Wollack}, \& {Zhao}}]{spergel2015}
{Spergel}, D., {Gehrels}, N., {Baltay}, C., {et~al.} 2015, arXiv e-prints, arXiv:1503.03757.
\newblock \doarXiv{1503.03757}

\bibitem[{{Spohn} {et~al.}(2022){Spohn}, {Savransky}, \& {Morgan}}]{spohn2022}
{Spohn}, C., {Savransky}, D., \& {Morgan}, R. 2022, \aj, 163, 163, \dodoi{10.3847/1538-3881/ac5049}

\bibitem[{{Stark} {et~al.}(2014){Stark}, {Roberge}, {Mandell}, \& {Robinson}}]{stark2014}
{Stark}, C.~C., {Roberge}, A., {Mandell}, A., \& {Robinson}, T.~D. 2014, \apj, 795, 122, \dodoi{10.1088/0004-637X/795/2/122}

\bibitem[{{The LUVOIR Team}(2019)}]{reportluvoir}
{The LUVOIR Team}. 2019, arXiv e-prints, arXiv:1912.06219.
\newblock \doarXiv{1912.06219}

\bibitem[{{Trifonov} {et~al.}(2020){Trifonov}, {Tal-Or}, {Zechmeister}, {Kaminski}, {Zucker}, \& {Mazeh}}]{trifonov2020}
{Trifonov}, T., {Tal-Or}, L., {Zechmeister}, M., {et~al.} 2020, \aap, 636, A74, \dodoi{10.1051/0004-6361/201936686}

\bibitem[{{Turnbull} {et~al.}(2021){Turnbull}, {Zimmerman}, {Girard}, {Hildebrandt}, {Li}, {Bogat}, {Gonzalez-Quiles}, {Stark}, {Mandell}, {Meshkat}, \& {Kane}}]{turnbull2021}
{Turnbull}, M.~C., {Zimmerman}, N., {Girard}, J.~H., {et~al.} 2021, Journal of Astronomical Telescopes, Instruments, and Systems, 7, 021218, \dodoi{10.1117/1.JATIS.7.2.021218}

\bibitem[{{Vaughan} {et~al.}(2023){Vaughan}, {Gebhard}, {Bott}, {Casewell}, {Cowan}, {Doelman}, {Kenworthy}, {Mazoyer}, {Millar-Blanchaer}, {Trees}, {Stam}, {Absil}, {Altinier}, {Baudoz}, {Belikov}, {Bidot}, {Birkby}, {Bonse}, {Brandl}, {Carlotti}, {Choquet}, {van Dam}, {Desai}, {Fogarty}, {Fowler}, {van Gorkom}, {Gutierrez}, {Guyon}, {Haffert}, {Herscovici-Schiller}, {Hours}, {Juanola-Parramon}, {Kleisioti}, {K{\"o}nig}, {van Kooten}, {Krasteva}, {Laginja}, {Landman}, {Leboulleux}, {Mouillet}, {N'Diaye}, {Por}, {Pueyo}, \& {Snik}}]{vaughan2023}
{Vaughan}, S.~R., {Gebhard}, T.~D., {Bott}, K., {et~al.} 2023, \mnras, 524, 5477, \dodoi{10.1093/mnras/stad2127}

\bibitem[{Virtanen {et~al.}(2020)Virtanen, Gommers, Oliphant, Haberland, Reddy, Cournapeau, Burovski, Peterson, Weckesser, Bright, {van der Walt}, Brett, Wilson, Millman, Mayorov, Nelson, Jones, Kern, Larson, Carey, Polat, Feng, Moore, {VanderPlas}, Laxalde, Perktold, Cimrman, Henriksen, Quintero, Harris, Archibald, Ribeiro, Pedregosa, {van Mulbregt}, \& {SciPy 1.0 Contributors}}]{virtanen2020}
Virtanen, P., Gommers, R., Oliphant, T.~E., {et~al.} 2020, Nature Methods, 17, 261, \dodoi{10.1038/s41592-019-0686-2}

\bibitem[{{W}es {M}c{K}inney(2010)}]{wes2010}
{W}es {M}c{K}inney. 2010, in {P}roceedings of the 9th {P}ython in {S}cience {C}onference, ed. {S}t\'efan van~der {W}alt \& {J}arrod {M}illman, 56 -- 61, \dodoi{10.25080/Majora-92bf1922-00a}

\bibitem[{{Wittenmyer} {et~al.}(2007){Wittenmyer}, {Endl}, \& {Cochran}}]{wittenmyer2007}
{Wittenmyer}, R.~A., {Endl}, M., \& {Cochran}, W.~D. 2007, \apj, 654, 625, \dodoi{10.1086/509110}

\bibitem[{{Wittenmyer} {et~al.}(2009){Wittenmyer}, {Endl}, {Cochran}, {Levison}, \& {Henry}}]{wittenmyer2009}
{Wittenmyer}, R.~A., {Endl}, M., {Cochran}, W.~D., {Levison}, H.~F., \& {Henry}, G.~W. 2009, \apjs, 182, 97, \dodoi{10.1088/0067-0049/182/1/97}

\bibitem[{{Wittenmyer} {et~al.}(2020){Wittenmyer}, {Wang}, {Horner}, {Butler}, {Tinney}, {Carter}, {Wright}, {Jones}, {Bailey}, {O'Toole}, \& {Johns}}]{wittenmyer2020b}
{Wittenmyer}, R.~A., {Wang}, S., {Horner}, J., {et~al.} 2020, \mnras, 492, 377, \dodoi{10.1093/mnras/stz3436}

\bibitem[{{Wolff} {et~al.}(2024){Wolff}, {Wang}, {Stapelfeldt}, {Bailey}, {Savransky}, {Hom}, {Biller}, {Brandner}, {Anche}, {Blunt}, {Brinjikji}, {Girard}, {Krause}, {Li}, {Livingston}, {Millar-Blanchaer}, {Noel}, {Pueyo}, {De Rosa}, {Samland}, \& {Schragal}}]{wolff2024}
{Wolff}, S.~G., {Wang}, J., {Stapelfeldt}, K., {et~al.} 2024, in Society of Photo-Optical Instrumentation Engineers (SPIE) Conference Series, Vol. 13092, Space Telescopes and Instrumentation 2024: Optical, Infrared, and Millimeter Wave, ed. L.~E. {Coyle}, S.~{Matsuura}, \& M.~D. {Perrin}, 1309255, \dodoi{10.1117/12.3020205}

\end{thebibliography}


\appendix

\section{Additional Figures}\label{appx:figures}

\begin{figure*}[htbp]
\includegraphics[width=\textwidth]{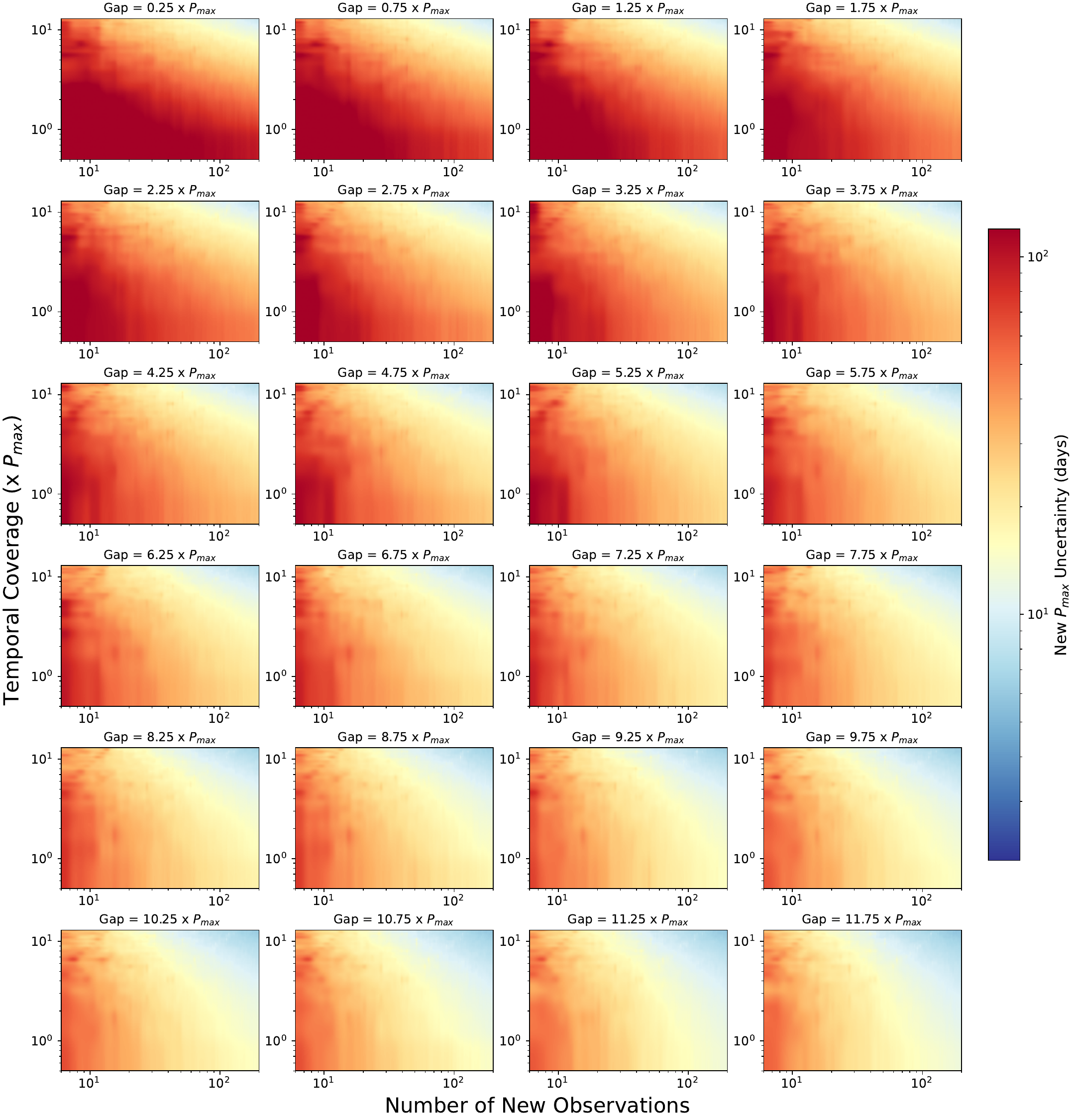}
\caption{Similar to Figure \ref{fig:rfmodel1}, except that the \cov~variable is now represented on the vertical axis and the \gap~variable is shown across the panels, with a step of 0.5 x\pmax.}
\label{fig:rfmodel2}
\end{figure*}

\begin{figure*}[htbp]
\includegraphics[width=\textwidth]{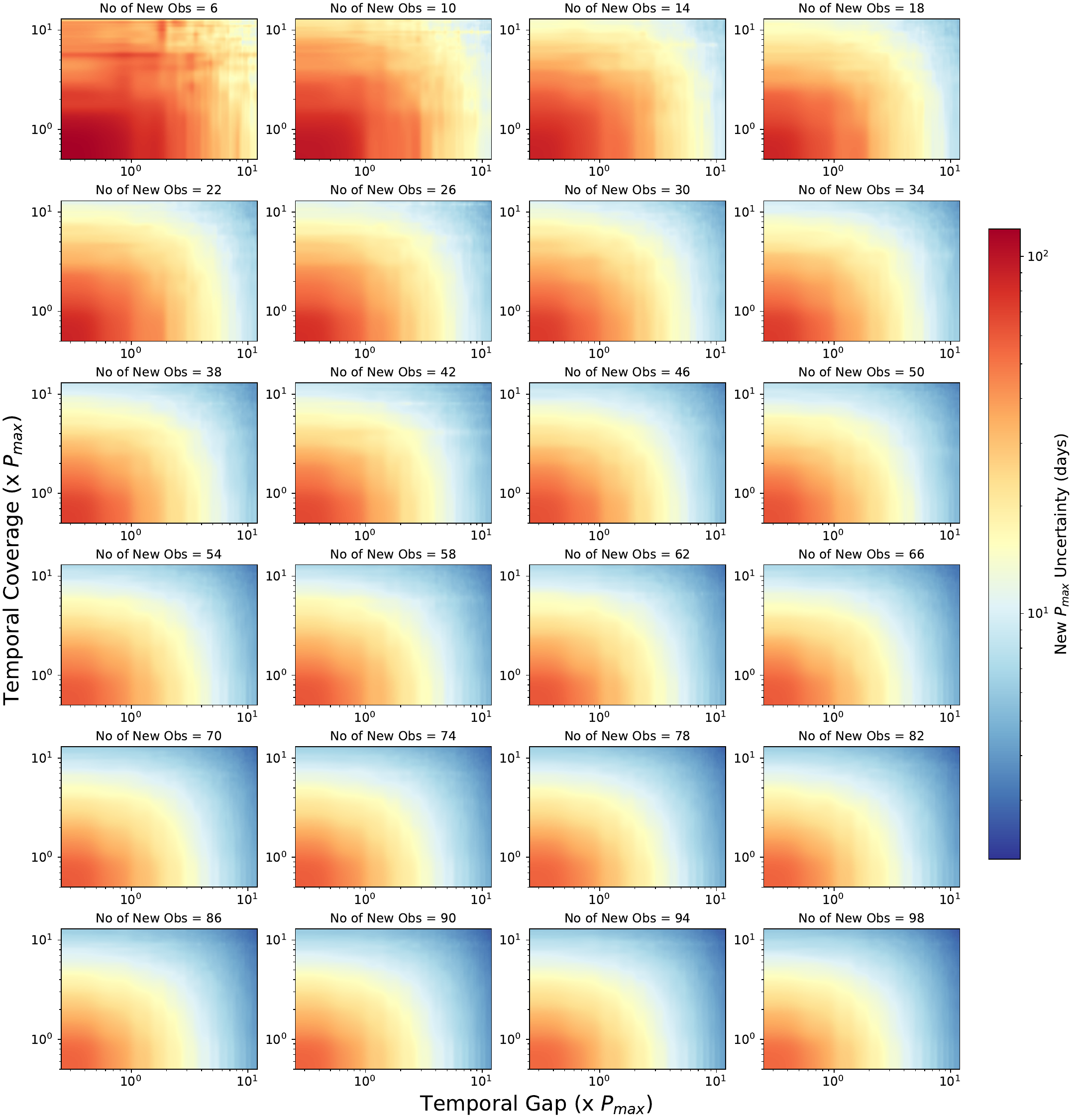}
\caption{Similar to Figure \ref{fig:rfmodel1}, except that the \gap~variable is now represented on the horizontal axis, the \cov~variable on the vertical axis, and the \num~variable is represented by the panels, with a step of 4 and up to \num~=~98. Higher values of \num~are not shown here as there is little change in the shape and color of the contour.}
\label{fig:rfmodel3}
\end{figure*}

\section{Additional Tables} \label{appx:tables}

\begin{deluxetable}{ccccc}[tbp]
    \tablecaption{Details of Short Simulation Runs
    \label{tab:rundetails}}
    \tablehead{
        \colhead{Target} & 
        \colhead{Run} &
        \colhead{Starting Time} &
        \colhead{\gap} &
        \colhead{\cov}
    }
    \startdata
     \multirow{8}{*}{\shortstack{HD 134987$^{a}$}} 
     & 1 & 2024-02-01 & 0.2561 & 0.2302 \\
     & 2 & 2024-08-01 & 0.2848 & 0.2015 \\
     & 3 & 2025-02-01 & 0.3138 & 0.1725 \\
     & 4	& 2025-08-01 & 0.3423 & 0.1440 \\
     & 5	& 2026-02-01 & 0.3713 & 0.1150 \\
     & 6	& 2026-08-01 & 0.3998 & 0.0865 \\
     & 7	& 2027-02-01 & 0.4288 & 0.0575 \\
     & 8	& 2027-08-01 & 0.4573 & 0.0290 \\
     \hline
     \multirow{8}{*}{\shortstack{47 UMa$^{b}$}}
     & 1	& 2024-02-01 & 0.0875 & 0.0898 \\
     & 2	& 2024-08-01 & 0.0987 & 0.0786 \\
     & 3	& 2025-02-01 & 0.1100 & 0.0673 \\
     & 4	& 2025-08-01 & 0.1212 & 0.0562 \\
     & 5	& 2026-02-01 & 0.1325 & 0.0449 \\
     & 6	& 2026-08-01 & 0.1436 & 0.0337 \\
     & 7	& 2027-02-01 & 0.1549 & 0.0224 \\
     & 8	& 2027-08-01 & 0.1660 & 0.0113 \\
     \hline
     \multirow{8}{*}{\shortstack{14 Her$^{c}$}}
     & 1	& 2024-02-01 & 0.0720 & 0.0732 \\
     & 2	& 2024-08-01 & 0.0811 &	0.0641 \\
     & 3	& 2025-02-01 & 0.0903 &	0.0549 \\
     & 4	& 2025-08-01 & 0.0994 &	0.0458 \\
     & 5	& 2026-02-01 & 0.1086 &	0.0366 \\
     & 6	& 2026-08-01 & 0.1177 &	0.0275 \\
     & 7	& 2027-02-01 & 0.1269 &	0.0183 \\
     & 8	& 2027-08-01 & 0.1360 &	0.0092 \\
    \enddata
    \tablecomments{The \gap~and the \cov~variables are in units of \pmax.}
    \tablenotetext{a}{Timestamp of last real RV observation at BJD 2,458,715.728759 (2019-08-20).}
    \tablenotetext{b}{Timestamp of last real RV observation at BJD 2,458,916.894692 (2020-03-08).}
    \tablenotetext{c}{Timestamp of last real RV observation at BJD 2,458,906.164844 (2020-02-26).}
\end{deluxetable}
    
\end{document}